\documentclass[%
reprint,
superscriptaddress,
amsmath,amssymb,
aps,
prd,
noeprint,
floatfix,
longbibliography
]{revtex4-1}
\pdfoutput=1
\usepackage{graphicx}
\usepackage{multirow}
\usepackage{pifont}
\usepackage{color}

\begin{document}
\title{Simulating the neutrino flux from the Spallation Neutron Source for the COHERENT experiment}
\date{\today}
\collaboration{COHERENT collaboration}
\newcommand{\Mephidesc}{\affiliation{National Research Nuclear University MEPhI (Moscow Engineering Physics Institute), Moscow, 115409, Russian Federation}}
\newcommand{\Dukedesc}{\affiliation{Department of Physics, Duke University, Durham, NC, 27708, USA}}
\newcommand{\TUNLdesc}{\affiliation{Triangle Universities Nuclear Laboratory, Durham, NC, 27708, USA}}
\newcommand{\UTKdesc}{\affiliation{Department of Physics and Astronomy, University of Tennessee, Knoxville, TN, 37996, USA}}
\newcommand{\ITEPdesc}{\affiliation{Institute for Theoretical and Experimental Physics named by A.I. Alikhanov of National Research Centre ``Kurchatov Institute'', Moscow, 117218, Russian Federation}}
\newcommand{\ORNLdesc}{\affiliation{Oak Ridge National Laboratory, Oak Ridge, TN, 37831, USA}}
\newcommand{\USDdesc}{\affiliation{Physics Department, University of South Dakota, Vermillion, SD, 57069, USA}}
\newcommand{\NCSUdesc}{\affiliation{Department of Physics, North Carolina State University, Raleigh, NC, 27695, USA}}
\newcommand{\Sandiadesc}{\affiliation{Sandia National Laboratories, Livermore, CA, 94550, USA}}
\newcommand{\UWdesc}{\affiliation{Center for Experimental Nuclear Physics and Astrophysics \& Department of Physics, University of Washington, Seattle, WA, 98195, USA}}
\newcommand{\LANLdesc}{\affiliation{Los Alamos National Laboratory, Los Alamos, NM, 87545, USA}}
\newcommand{\Laurentiandesc}{\affiliation{Department of Physics, Laurentian University, Sudbury, Ontario, P3E 2C6, Canada}}
\newcommand{\CMUdesc}{\affiliation{Department of Physics, Carnegie Mellon University, Pittsburgh, PA, 15213, USA}}
\newcommand{\IUdesc}{\affiliation{Department of Physics, Indiana University, Bloomington, IN, 47405, USA}}
\newcommand{\VTdesc}{\affiliation{Center for Neutrino Physics, Virginia Tech, Blacksburg, VA, 24061, USA}}
\newcommand{\NCCUdesc}{\affiliation{Department of Mathematics and Physics, North Carolina Central University, Durham, NC, 27707, USA}}
\newcommand{\UFdesc}{\affiliation{Department of Physics, University of Florida, Gainesville, FL, 32611, USA}}
\newcommand{\Tuftsdesc}{\affiliation{Department of Physics and Astronomy, Tufts University, Medford, MA, 02155, USA}}
\newcommand{\SNUdesc}{\affiliation{Department of Physics and Astronomy, Seoul National University, Seoul, 08826, Korea}}
\author{D.~Akimov}\Mephidesc
\author{P.~An}\Dukedesc\TUNLdesc
\author{C.~Awe}\Dukedesc\TUNLdesc
\author{P.S.~Barbeau}\Dukedesc\TUNLdesc
\author{B.~Becker}\UTKdesc
\author{V.~Belov }\ITEPdesc\Mephidesc
\author{I.~Bernardi}\UTKdesc
\author{M.A.~Blackston}\ORNLdesc
\author{C.~Bock}\USDdesc
\author{A.~Bolozdynya}\Mephidesc
\author{J.~Browning}\NCSUdesc
\author{B.~Cabrera-Palmer}\Sandiadesc
\author{D.~Chernyak}\altaffiliation{Now at: Department of Physics and Astronomy, University of Alabama, Tuscaloosa, AL, 35487, USA and Institute for Nuclear Research of NASU, Kyiv, 03028, Ukraine}\USDdesc
\author{E.~Conley}\Dukedesc
\author{J.~Daughhetee}\ORNLdesc
\author{J.~Detwiler}\UWdesc
\author{K.~Ding}\USDdesc
\author{M.R.~Durand}\UWdesc
\author{Y.~Efremenko}\UTKdesc\ORNLdesc
\author{S.R.~Elliott}\LANLdesc
\author{L.~Fabris}\ORNLdesc
\author{M.~Febbraro}\ORNLdesc
\author{J.~Galambos}\ORNLdesc
\author{A.~Gallo Rosso}\Laurentiandesc
\author{A.~Galindo-Uribarri}\ORNLdesc\UTKdesc
\author{M.P.~Green }\TUNLdesc\ORNLdesc\NCSUdesc
\author{M.R.~Heath}\ORNLdesc
\author{S.~Hedges}\Dukedesc\TUNLdesc
\author{D.~Hoang}\CMUdesc
\author{M.~Hughes}\IUdesc
\author{E.~Iverson}\ORNLdesc
\author{T.~Johnson}\Dukedesc\TUNLdesc
\author{A.~Khromov}\Mephidesc
\author{A.~Konovalov}\Mephidesc\ITEPdesc
\author{E.~Kozlova}\Mephidesc\ITEPdesc
\author{A.~Kumpan}\Mephidesc
\author{L.~Li}\Dukedesc\TUNLdesc
\author{J.M.~Link}\VTdesc
\author{J.~Liu}\USDdesc
\author{K.~Mann}\NCSUdesc
\author{D.M.~Markoff}\NCCUdesc\TUNLdesc
\author{J.~Mastroberti}\IUdesc
\author{M.~McIntyre}\UFdesc
\author{P.E.~Mueller}\ORNLdesc
\author{J.~Newby}\ORNLdesc
\author{D.S.~Parno}\CMUdesc
\author{S.I.~Penttila}\ORNLdesc
\author{D.~Pershey}\Dukedesc
\author{R.~Rapp}\email{rrapp@andrew.cmu.edu}\CMUdesc
\author{H.~Ray}\UFdesc
\author{J.~Raybern}\Dukedesc
\author{O.~Razuvaeva}\Mephidesc\ITEPdesc
\author{D.~Reyna}\Sandiadesc
\author{G.C.~Rich}\TUNLdesc
\author{D.~Rimal}\UFdesc
\author{J.~Ross}\NCCUdesc\TUNLdesc
\author{D.~Rudik}\Mephidesc
\author{J.~Runge}\Dukedesc\TUNLdesc
\author{D.J.~Salvat}\IUdesc
\author{A.M.~Salyapongse}\CMUdesc
\author{K.~Scholberg}\Dukedesc
\author{A.~Shakirov}\Mephidesc
\author{G.~Simakov}\Mephidesc\ITEPdesc
\author{G.~Sinev}\altaffiliation{Now at: South Dakota School of Mines and Technology, Rapid City, SD, 57701, USA}\Dukedesc
\author{W.M.~Snow}\IUdesc
\author{V.~Sosnovstsev}\Mephidesc
\author{B.~Suh}\IUdesc
\author{R.~Tayloe}\IUdesc
\author{K.~Tellez-Giron-Flores}\VTdesc
\author{I.~Tolstukhin}\altaffiliation{Now at: Argonne National Laboratory, Argonne, IL, 60439, USA}\IUdesc
\author{S.~Trotter}\ORNLdesc
\author{E.~Ujah}\NCCUdesc\TUNLdesc
\author{J.~Vanderwerp}\IUdesc
\author{R.L.~Varner}\ORNLdesc
\author{C.J.~Virtue}\Laurentiandesc
\author{G.~Visser}\IUdesc
\author{T.~Wongjirad}\Tuftsdesc
\author{Y.-R.~Yen}\CMUdesc
\author{J.~Yoo}\SNUdesc
\author{C.-H.~Yu}\ORNLdesc
\author{J.~Zettlemoyer}\altaffiliation{Now at: Fermi National Accelerator Laboratory, Batavia, IL, 60510, USA}\IUdesc
\author{S.~Zhang}\CMUdesc\IUdesc

\begin{abstract}
  The Spallation Neutron Source (SNS) at Oak Ridge National Laboratory
  is a pulsed source of neutrons and, as a byproduct of this
  operation, an intense source of pulsed neutrinos via stopped-pion
  decay.  The COHERENT collaboration uses this source to investigate
  coherent elastic neutrino-nucleus scattering and other physics with
  a suite of detectors.  This work includes a description of our
  Geant4 simulation of neutrino production at the SNS and the flux
  calculation which informs the COHERENT studies.  We estimate the
  uncertainty of this calculation at $\sim$~10\% based on validation
  against available low-energy $\pi^+$ production data.
\end{abstract}

\maketitle
\section{Introduction}

At the Spallation Neutron Source (SNS) at Oak Ridge National
Laboratory (ORNL), a pulsed 1.4-MW beam of $\sim$~1-GeV protons
strikes an approximately 50~cm-long Hg target \cite{sns_design}.  The
incident protons interact multiple times within the thick target,
losing energy and spalling nuclei to create the intended neutrons and
byproduct charged pions.  The majority of the $\pi^+$ come to rest
(less than 1\% decay in flight) within the thick and dense target, and
their stopped decays then give rise to neutrinos with energies of
order tens of MeV:
\begin{equation}
\begin{aligned}
  \pi^+ \rightarrow \hspace{2pt}&\mu^+ + \nu_\mu\\
  &\mu^+ \rightarrow e^+ + \bar{\nu}_\mu + \nu_e .
\end{aligned}
\end{equation}
\noindent SNS interactions also produce copious quantities of $\pi^-$,
but the vast majority ($\sim$~99\%) of these capture on nuclei in the
target before decaying and rarely produce neutrinos. The proton beam
energy is too low to create substantial numbers of other
neutrino-producing decay chains, such as those of $K^{\pm}$ or $\eta$.

To take advantage of this high-intensity pulsed-neutrino source, the
COHERENT collaboration has deployed multiple neutrino detectors
20-30~m from the target in the SNS basement corridor known as
``Neutrino Alley.'' The collaboration has performed the first-ever
measurements of the cross section of coherent elastic neutrino-nucleus
scattering (CEvNS) with the COH-CsI~\cite{sci_mag} and COH-Ar-10
detectors~\cite{coherent-Ar:2021}. CEvNS measurements are planned on
additional nuclear targets (Na and Ge), as well as measurements of
several charged-current neutrino-interaction cross sections of
interest to nuclear and particle physics and
astrophysics~\cite{arxiv2018}. As COHERENT's cross-section
measurements become more precise, they will illuminate physics topics
including non-standard neutrino interactions~\cite{barranco:2005,
  deNiverville-LightNewPhysics:2015}, neutrino electromagnetic
properties~\cite{VogelEngel:1989, Scholberg:2005qs,
  Papavassiliou:2005cs, Kosmas:2015, Cadeddu-ChargeRadius:2018},
nuclear form factors and neutron distributions~\cite{Amanik_2008,
  Patton-NeutronDensity-2012, Caddedu-CsINeutronDensity-2018}, and the
detection of supernova neutrinos by both dedicated
observatories~\cite{duba:2008, vaananen:2011} and next-generation
neutrino-oscillation
experiments~\cite{Langanke-SupernovaNu-WaterCherenkov:1996,
  Scholberg:2012id, duneSNB}.

Precise knowledge of the SNS neutrino flux is essential to unlocking
the full physics potential of the COHERENT cross-section
measurements. The error on the overall normalization of the neutrino
flux is the dominant systematic in the Ar
results~\cite{coherent-Ar:2021} and the second-largest systematic in
the initial CsI results~\cite{sci_mag}.  Thanks to updated
measurements of the quenching factor in CsI, the neutrino flux is the
dominant systematic in the final CsI results~\cite{csi2020}.

We have built a detailed model of the SNS using the Geant4 Monte Carlo
framework~\cite{geant4, Allison:2006ve} to characterize the neutrino
flux to the COHERENT detectors and have made it publicly available on
Zenodo \cite{snsFluxSimsRelease}.  In addition to the geometry, the
simulation accuracy relies on the underlying implementation of pion
production in the Geant4 physics model.  Section~\ref{sec:physlists}
describes our validation efforts using four standard physics lists
against the available world $\pi^+$-production data.  World data,
however, are imperfect --- Hg-target data are not available at low
proton energies, data sets at proton energies near 1~GeV are very
limited, and most pion-production cross sections are measured using
thin targets that do not replicate the half-meter of dense material
the protons at the SNS encounter.  Although the existing data are
insufficient for a precise validation, we estimate the uncertainty of
our simulated flux with the QGSP\_\hspace{2pt}BERT physics list to be
about 10\%.  Section~\ref{sec:geant4} describes our simulation of the
SNS, along with our tools for studying the characteristics of the
resulting neutrinos.  We also discuss the effect of changes to SNS
operating conditions; for example, the incident proton kinetic energy
has ranged from 0.83 -- 1.011~GeV during COHERENT's lifetime in
Neutrino Alley.  Section~\ref{sec:fts} summarizes the properties of
our simulated neutrino flux using the selected physics list.

Our SNS simulation has applications to additional nuclear and particle
physics experiments proposed at the SNS. In Section~\ref{sec:sts}, we
present a neutrino-flux simulation based on preliminary design work
for a proposed Second Target Station (STS) with a tungsten target; our
results suggest that the STS could be a very productive site for
next-generation neutrino experiments. Section~\ref{sec:darkmatter}
describes the use of our simulation to study $\pi^0$ and $\pi^-$
production at the SNS, relevant to accelerator-based searches for
light dark matter.  We discuss several future avenues for reducing
uncertainties related to the SNS neutrino flux in
Section~\ref{sec:future} and conclude in Section~\ref{sec:theEnd}.

\section{Validation of Simulation Physics}
\label{sec:physlists}

We investigated four standard physics models (or ``physics lists'') as
implemented in Geant4.10.06.p01: FTFP\_\hspace{2pt}BERT,
QGSP\_\hspace{2pt}BERT, QGSP\_\hspace{2pt}BIC, and
QGSP\_\hspace{2pt}INCLXX.  With all SNS protons well below 10~GeV, the
differences in the underlying string models of FTFP\_\hspace{2pt}BERT
and QGSP\_\hspace{2pt}BERT were found to be negligible; in this work
we focus only on QGSP\_\hspace{2pt}BERT.  We note here that the plots
within this section use natural units, such that $c = 1$.

\par\indent Each candidate for our physics list models nuclear
structure in a specific way.  With an implementation of the classical
Bertini Cascade model \cite{bertini} for incident hadrons below 3~GeV,
QGSP\_\hspace{2pt}BERT is a favored model for the production of
hadrons (and subsequently, neutrinos) with its treatment of the
nucleus as a gas of nucleons that can be solved on average using the
Boltzmann equation for a projectile moving through the gas
\cite{g4bert}.  The QGSP\_\hspace{2pt}BIC physics list differs only
for protons and neutrons, for which it implements a Binary Cascade and
models the nucleus as an isotropic sphere.  In this model, the
nucleons are placed at specific positions that projectiles can
interact with individually, and each nucleon carries a random momentum
between zero and the Fermi momentum \cite{g4binary}.  Finally,
QGSP\_\hspace{2pt}INCLXX extends the Liege Intranuclear Cascade model
\cite{liege} benchmarked against spallation studies below 3~GeV
\cite{inclxxSpalTest} by modeling the nucleus in a very similar manner
to QGSP\_\hspace{2pt}BIC, but adding the possibility to emit nucleon
clusters that can cause secondary reactions after a projectile
interacts with the nucleus.  Both QGSP\_\hspace{2pt}BIC and
QGSP\_\hspace{2pt}INCLXX require increased computation time to model
the interactions of projectiles with more massive nuclei (compared to
QGSP\_\hspace{2pt}BERT) \cite{g4inclxx}.

\par\indent In prior estimations the COHERENT collaboration has used
the QGSP\_\hspace{2pt}BERT physics list with an assigned 10\%
uncertainty on any flux predictions coming from simulation efforts.
This estimate was informed by prior studies using an implementation of
the Bertini model in the LAHET Monte Carlo framework \cite{lahet} to
make predictions for the LSND and KARMEN experiments \cite{lahetLSND,
  lahetKARMEN, lahetKARMEN2}.  World data at the time of their
investigation did not agree with LAHET predictions, and LAHET
predictions that were renormalized to match available data were lower
than Geant4 predictions \cite{lahetLSND, lahetData1, lahetData2}.  The
10\% systematic was assigned to our neutrino flux calculations to
conservatively account for this discrepancy~\cite{sci_mag}.

\par Since the lack of pion-production data from 1~GeV proton-mercury
interactions prevents a direct comparison, our choice of physics model
must be validated via other targets, usually at higher energies.  In
Section~\ref{sec:nt} we compare the total $\pi^+$-production cross
section to the Norbury-Townsend parameterization developed to match
data from proton-nucleus and nucleus-nucleus
collisions~\cite{ntParameterization}.  Not included in the development
of the Norbury-Townsend parameterization, however, are newer results
focusing on double-differential measurements, such as those from the
thin-target HARP experiment~\cite{Catanesi:2007ig}.  We detail our
validations against the HARP measurements in Section~\ref{sec:harp}.
Older experiments also collected double-differential pion-production
data at energies closer to the SNS, such as Abaev \textit{et al.}~in
1989~\cite{Abaev:1988az}, but their data have a very limited angular
coverage.  We use these data sets to check the model behavior at lower
proton energies (Section~\ref{sec:abaev}) since they cannot constrain
our total neutrino flux.  We discuss the effects of modeling the thick
target of the SNS in Section~\ref{sec:thick} and interpret all of our
validation work to estimate a neutrino flux systematic for COHERENT in
Section~\ref{sec:sys}.

\subsection{Norbury-Townsend Parameterization}
\label{sec:nt}
\begin{figure}
  \centering
  \includegraphics[width = \columnwidth]{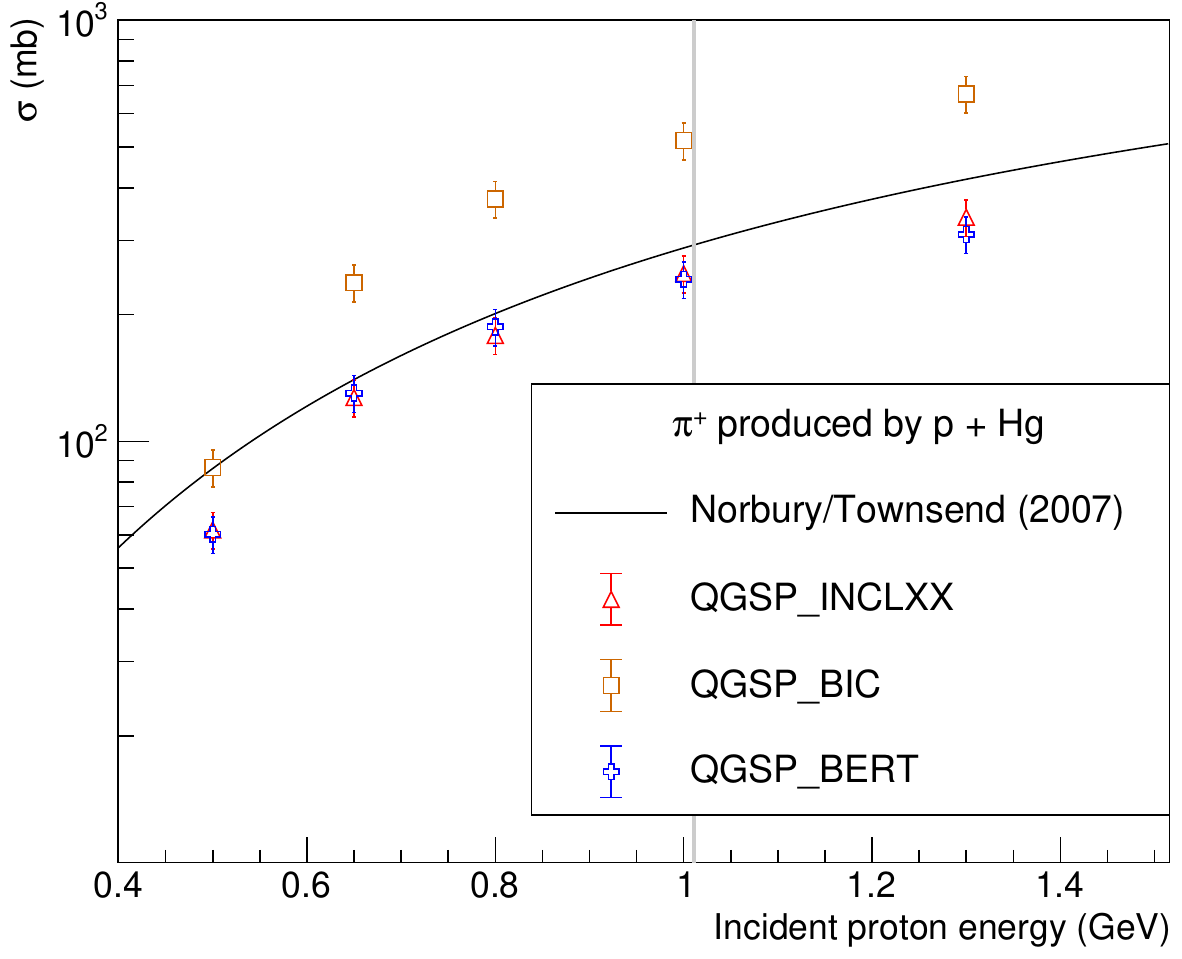}
  \includegraphics[width = \columnwidth]{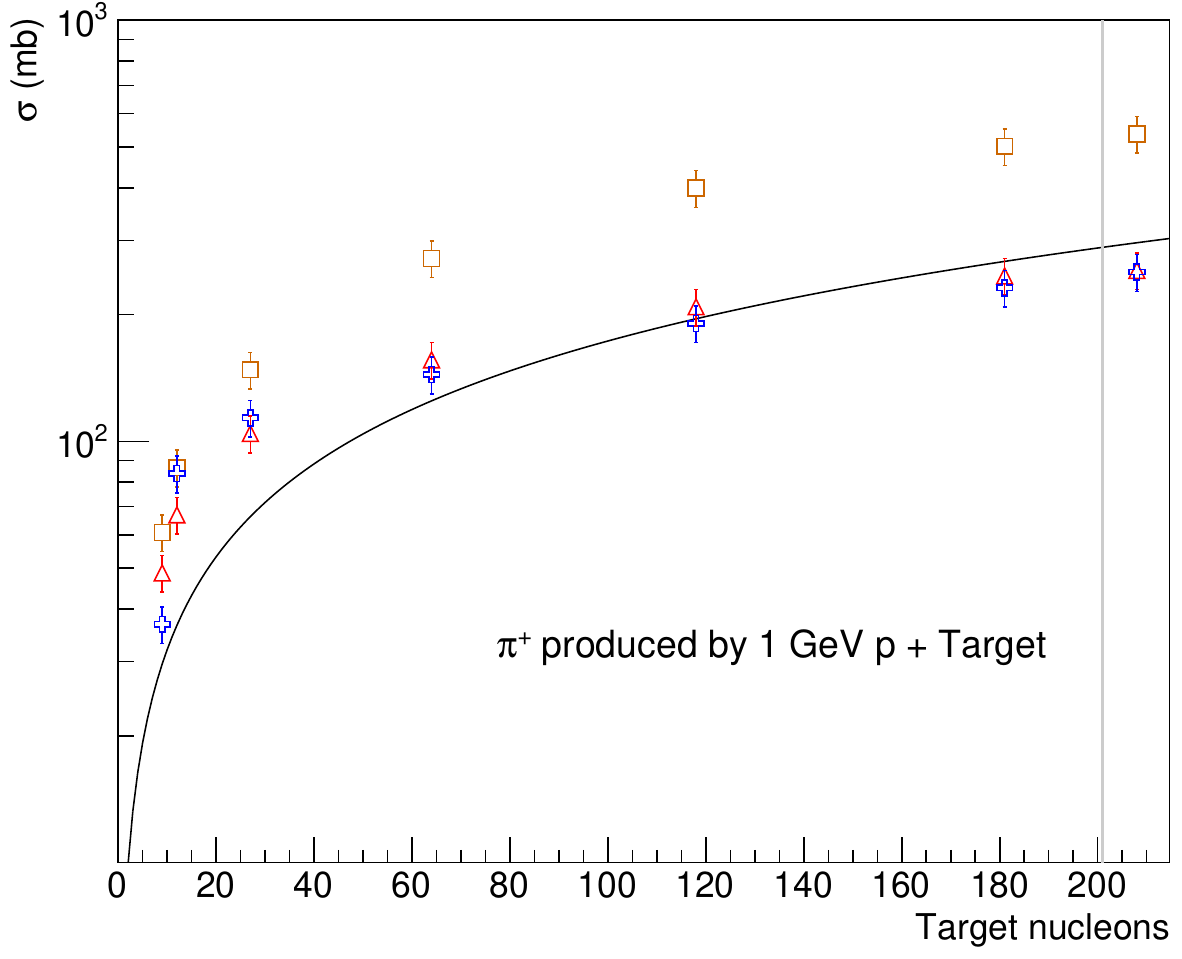}
  \caption{Comparisons of the Norbury-Townsend parameterization and
    Geant4 model predictions of total pion-production cross section.
    Top: Dependence of total cross section on incident proton energy
    for a mercury target.  The vertical line indicates the current SNS
    operating energy of 1.011~GeV, but COHERENT still sees $\pi^+$
    production at energies below this value due to proton energy loss
    within the thick target (see Fig.~\ref{fig:pe}).  Bottom:
    Dependence of total cross section on target nucleus for a proton
    energy of 1~GeV.  The vertical line represents a mercury target.}
  \label{fig:N/T}
\end{figure}

This empirical function was developed to parametrize pion-production
data from proton-nucleus and nucleus-nucleus interactions measured by
Nagamiya \textit{et al.} \cite{ntData}.  While developed in the right
energy range for SNS operations at $\sim$~1~GeV, $\pi^+$ production
data was only taken for subsets of Ne + NaF, Ne + Cu, Ne + Pb, C + C,
C + Pb, Ar + KCl, Ar + Pb for 0.4, 0.8, and 2.1~GeV per incident
nucleon --- only $\pi^-$ production data was available from the
proton-nucleus studies~\cite{ntParameterization}.  Although our focus
is $\pi^+$ production in this work, we note that future effort to
check the candidate physics models against the parameterizations for
$\pi^-$ and $\pi^0$ production will be useful to validate the flux
predictions for dark-matter-producing particles at the SNS that we
present in Section \ref{sec:darkmatter}.

The Norbury-Townsend parameterization of the $\pi^+$ production cross
section ($\sigma_{\pi^+}$ in mb) is shown in Eqn.~\ref{eq:NT}, where
$A_t$ is the number of target nucleons, and $E_i$ (in GeV) is the
energy per incident nucleon:
\begin{equation}
  \sigma_{\pi^+} = \frac{A_t^{2.2/3}}{0.00717 + 0.0652\frac{\log\left(E_i\right)}{E_i} +
    \frac{0.162}{E_i^2}}.
  \label{eq:NT}
\end{equation}

\par Using a thin simulated target (5 $\times$ 5 $\times$ 0.5~cm$^3$)
with specified isotope, molar mass, and density, we counted the total
number of pions produced.  We then scaled this event rate by our
simulated number of target nuclei and incident flux of protons to
convert to a total cross section prediction.  Figure \ref{fig:N/T}
shows comparisons of these results to the parameterization across
incident energies (top) and target nucleus (bottom), with a 10\%
uncertainty applied to the cross sections from each potential physics
list.

\subsection{HARP and HARP-CDP}
\label{sec:harp}

\begin{figure*}
  \includegraphics[width = \textwidth]{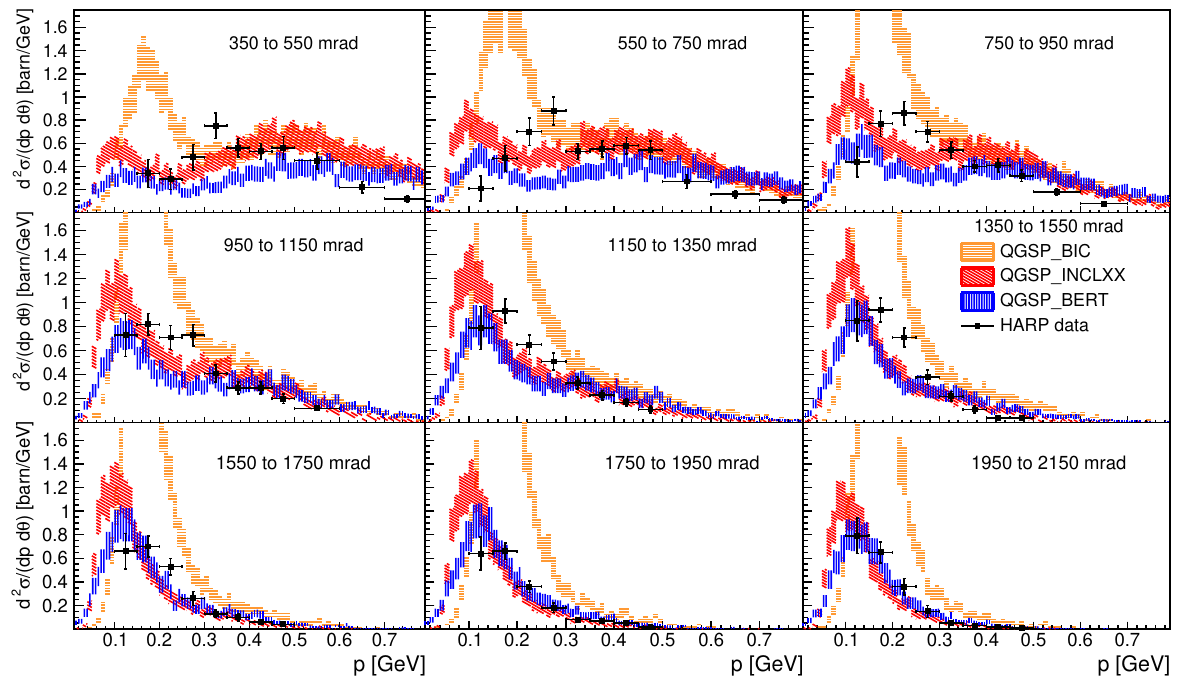}  
  \includegraphics[width = \textwidth]{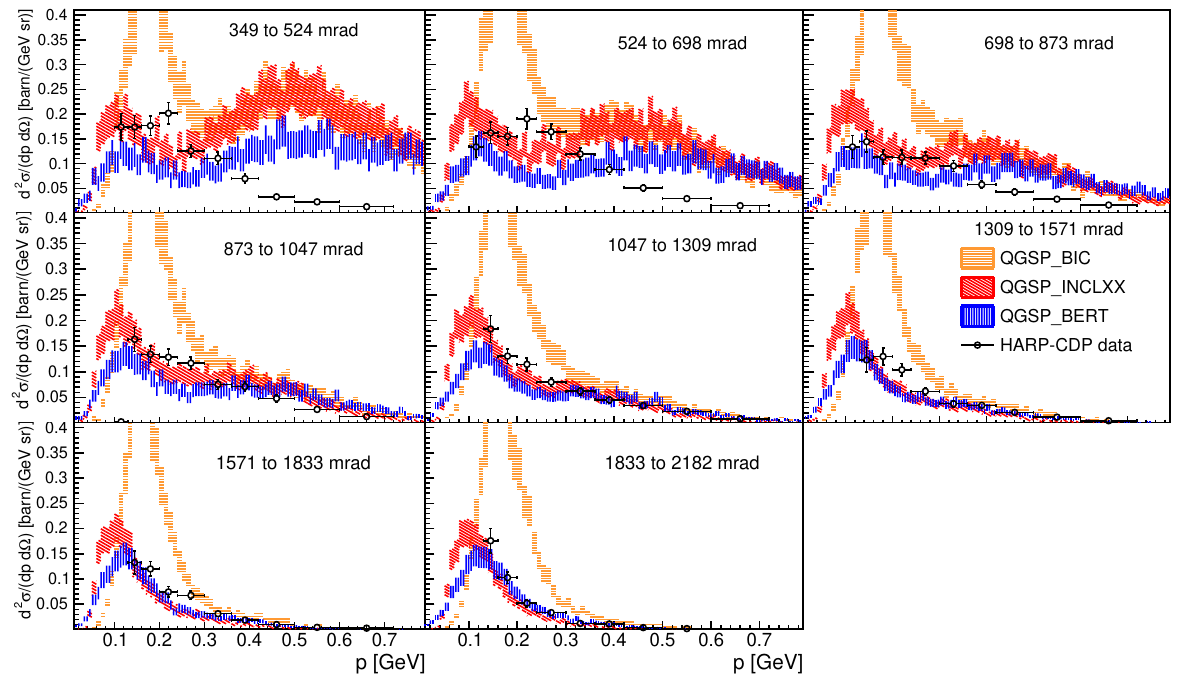}
  \caption{Comparisons of double-differential cross sections of
    $\pi^+$ production from 3~GeV/c p$ + ^{208}$Pb as predicted by the
    different Geant4 physics lists to the measurements from HARP (top)
    and HARP-CDP (bottom).  The error band shown for each physics list
    is a 10\% uncertainty.  We highlight $^{208}$Pb because, among
    HARP targets, this isotope is closest in mass to the SNS mercury
    target.}
  \label{fig:doubleDiffs}
\end{figure*}

HARP, the Hadron Production Experiment (PS214), operated at CERN's
Proton Synchrotron from 2000 to 2002. With a nearly 4$\pi$ acceptance
and incident proton momentum range from 1.5 GeV/c to 15 GeV/c, HARP
measured 7 different solid targets (Be, C, Al, Cu, Sn, Ta, Pb) as well
as 4 cryogenic liquid targets (H$_2$, D$_2$, N$_2$, O$_2$).  The HARP
collaboration disagreed on their TPC calibrations causing a subgroup,
HARP-CDP, to promote different calculations of pion momenta and
identification of protons and pions \cite{HARP-CDP}.  Both of these
differences impact the final analysis, such that HARP-CDP reports
lower cross sections than the HARP analysis. In this paper, both sets
of cross section results were checked against our Monte Carlo
simulations.

Data were not collected for incident protons at 1~GeV; therefore we
compare to the HARP and HARP-CDP analyses of 3~GeV/c data on a large
range of nucleon numbers: Be \cite{beAlPbData}, C \cite{cCuSnData,
  cCDPdata}, Al \cite{beAlPbData, alCDPdata}, Cu \cite{cCuSnData}, Sn
\cite{cCuSnData, snCDPdata}, Ta \cite{taData, taCDPdata}, and Pb
\cite{beAlPbData, pbCDPdata}.  We follow a similar procedure to our
Norbury-Townsend comparisons and simulate monoenergetic protons with
2.205~GeV of kinetic energy (calculated from the 3~GeV/c beam
momentum) incident on a thin target (5 $\times$ 5 $\times$
0.5~cm$^3$), though here counting pions produced per pion momentum and
production angle rather than total number of pions.  We then scale the
stored event rates by our simulated target details to convert to a
doubly differential cross section prediction from each simulation
model.  Figure \ref{fig:doubleDiffs} illustrates the direct comparison
of our simulation to the HARP and HARP-CDP results for 3~GeV/c p $+$
$^{208}$Pb.  The simulation error bands combine statistical
uncertainty with the estimated 10\% systematic uncertainty on the
simulation prediction.

\begin{figure}
  \includegraphics[width = \columnwidth]{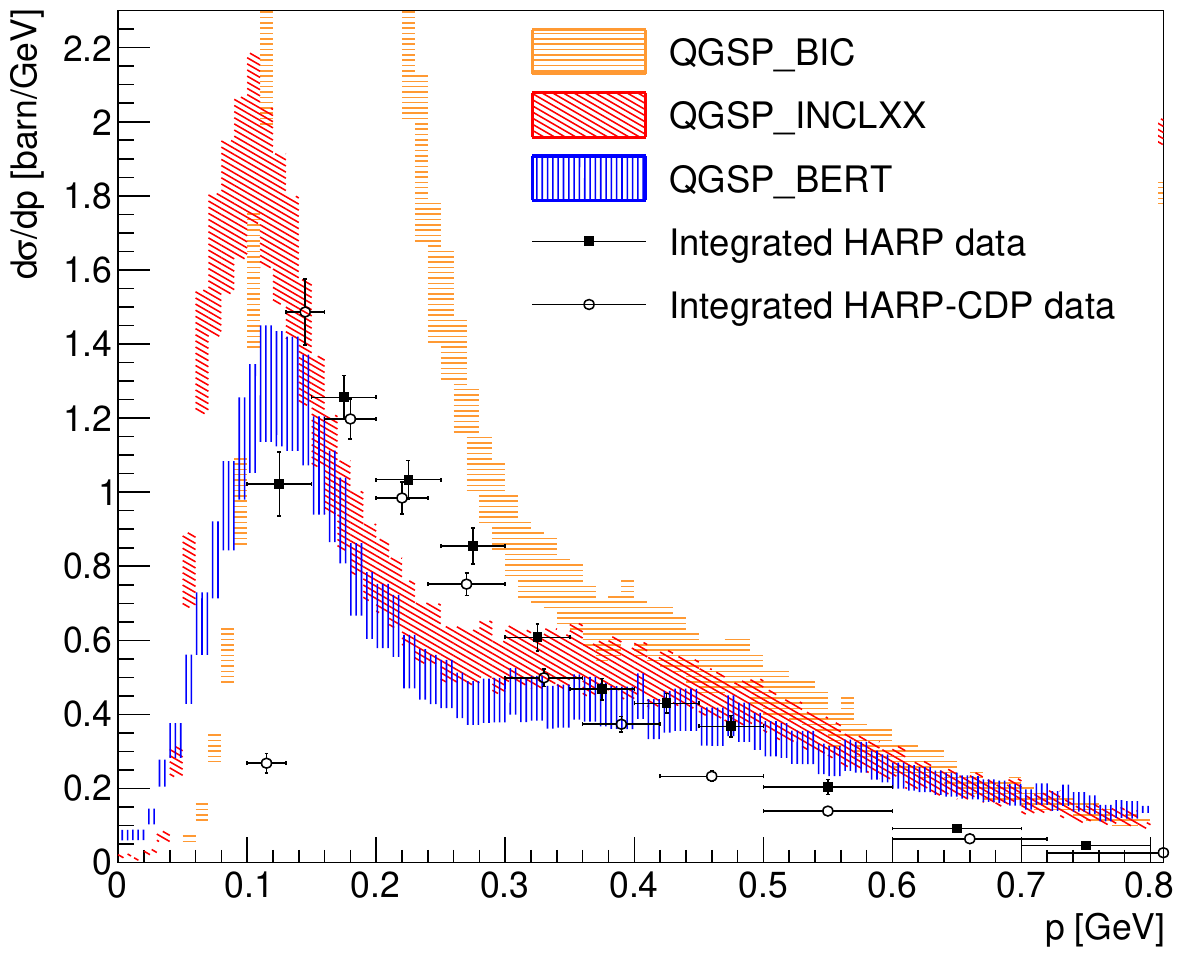}
  \includegraphics[width = \columnwidth]{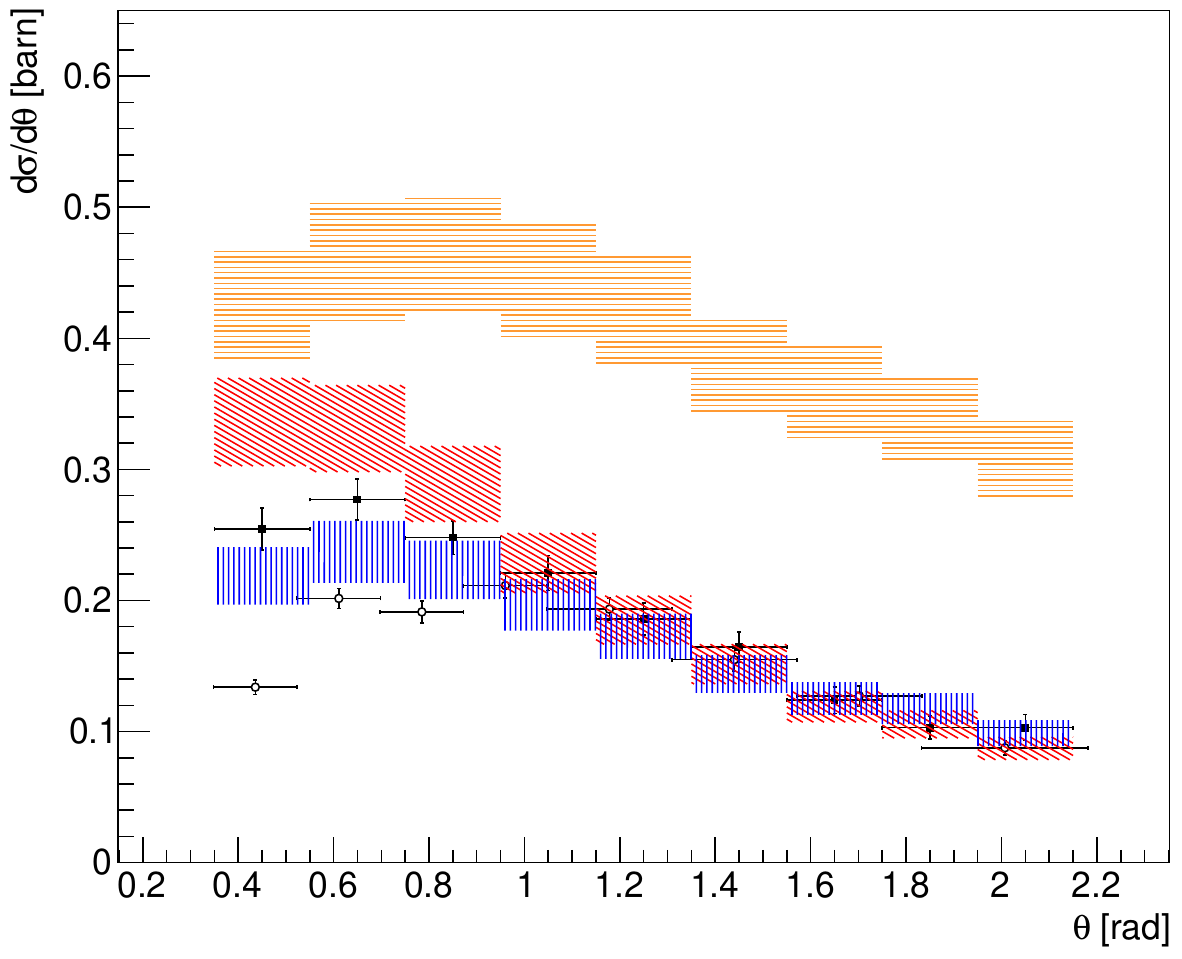}
  \caption{Comparison of measured differential cross sections of
    $\pi^+$ production from 3~GeV/c p$ + ^{208}$Pb to Geant4 physics
    lists.  Top: HARP and HARP-CDP data were integrated over their
    respective angular regions and compared to simulation integrated
    from 350 to 2150 mrad in production angle.  Bottom: HARP and
    HARP-CDP data were integrated from 0.1 to 0.8 GeV/c in momentum
    and compared to simulation integrated on the same region.}
  \label{fig:singleDiffs}
\end{figure}

\begin{figure}
  \includegraphics[width = \columnwidth]{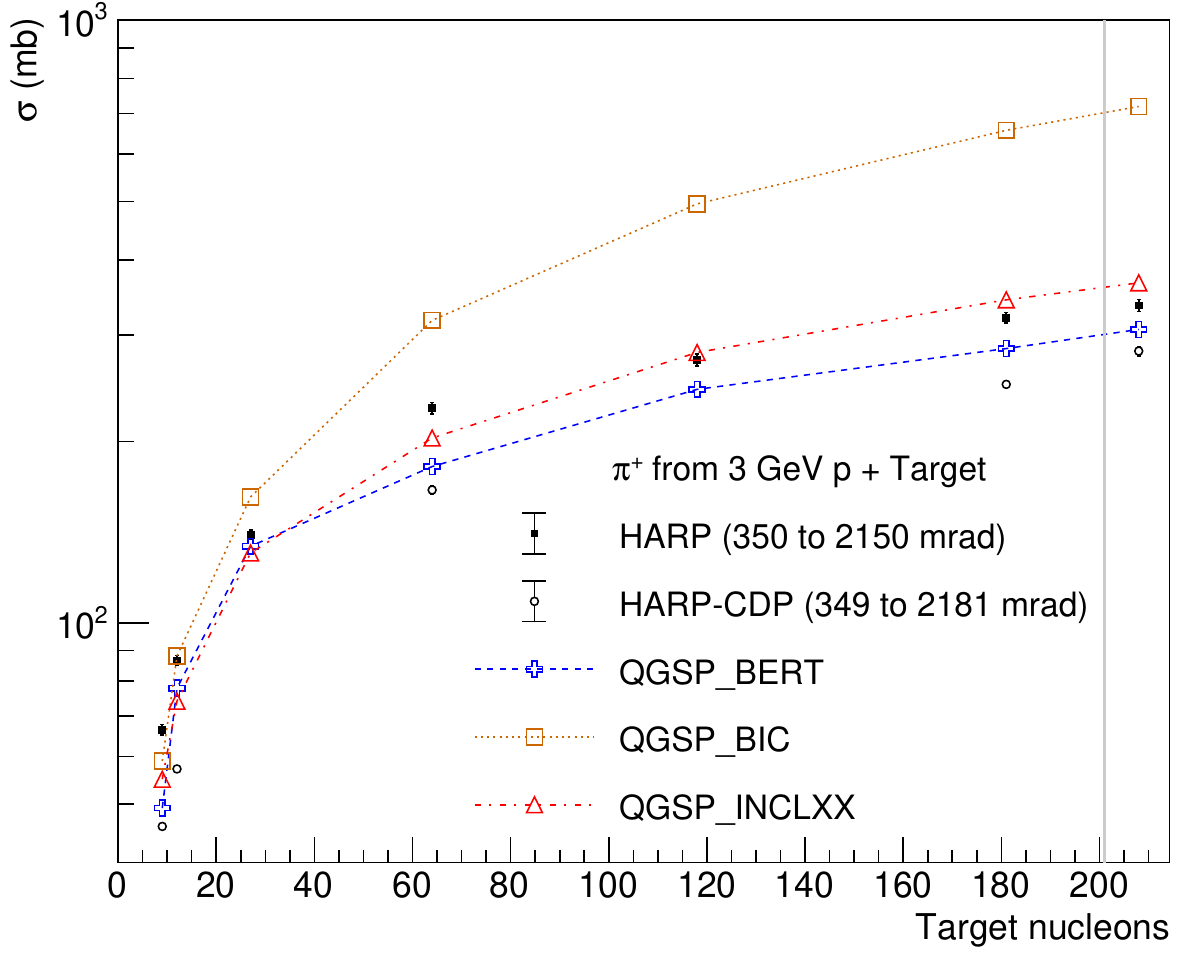}
  \includegraphics[width = \columnwidth]{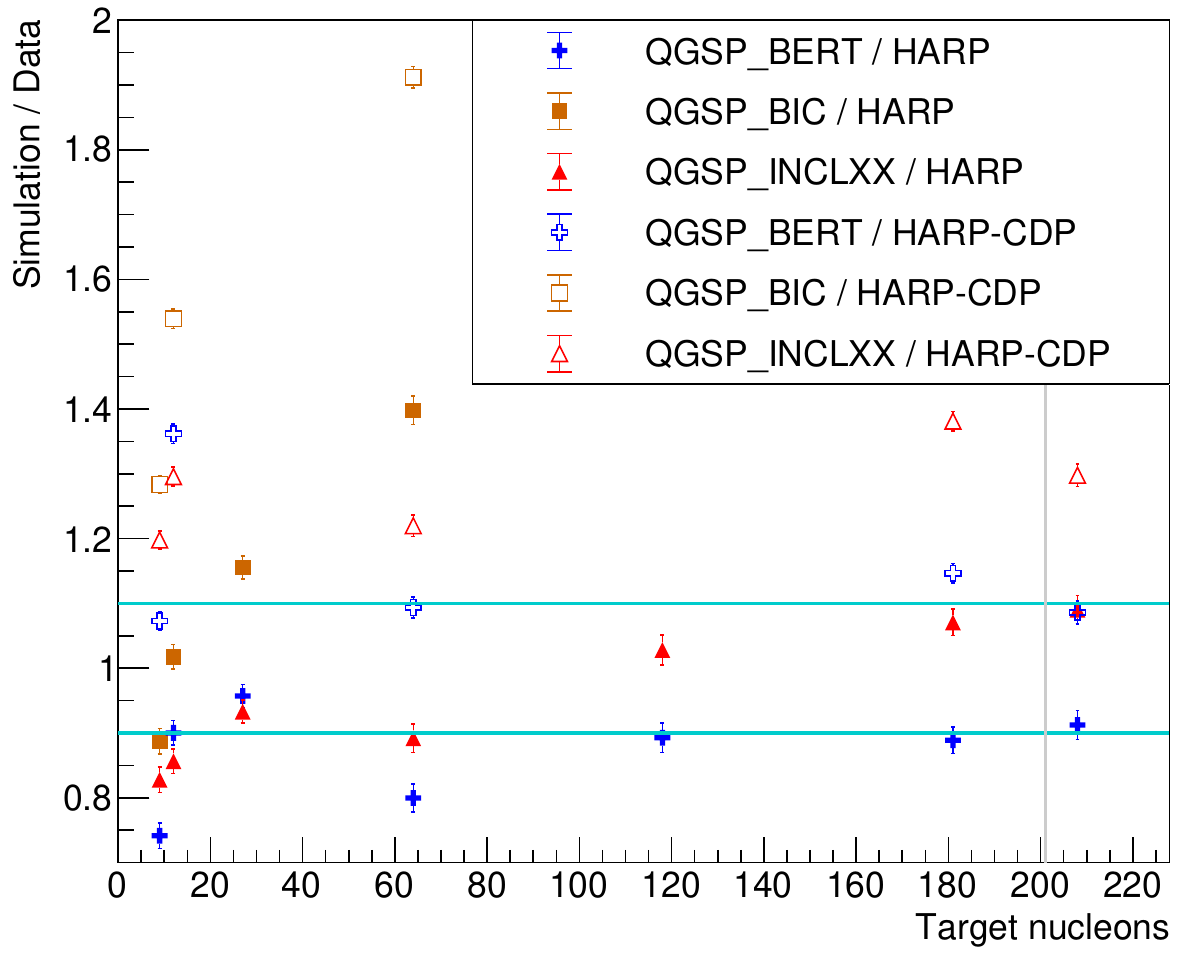}
  \caption{Top: The HARP data and Geant4 model predictions of the
    pion-production cross section integrated over 350 - 2150 mrad and
    0.1 - 0.8 GeV.  The HARP-CDP data are also shown but are
    integrated over 349 - 2181 mrad and 0.1 - 0.8 GeV.  Bottom: Ratio
    of the Geant4 simulated predictions to the central values of the
    data, plotted with an uncertainty on all three simulations shown
    as data error / central value (the HARP-CDP error bars are small
    enough to be hidden by the points themselves).  The horizontal
    cyan lines mark a $\pm$10\% uncertainty band.  The vertical gray
    line on each plot represents a mercury target.}
  \label{fig:HARPtotal}
\end{figure}

\par Since the Geant4 models predict the HARP and HARP-CDP data better
in some bins than others, we integrate away the angular or momentum
dependence to compare singly differential cross sections.  The
comparisons shown in Fig.~\ref{fig:singleDiffs} integrate our
simulation prediction over the angular region of the HARP analysis;
there is less than a 1\% difference from the simulation prediction
integrated over the HARP angular region and over the HARP-CDP angular
region, so we show HARP-CDP data on the same axes.

\par The SNS mercury target is thick and dense enough to stop the
majority of the pions regardless of production angle or momentum.
Therefore, to a very good approximation, we require only the total
cross section to simulate the neutrino production.  We integrate away
the dependence on both production angle (350 - 2150 mrad) and momentum
(0.1 - 0.8 GeV) and show the total cross-section comparisons to both
HARP and HARP-CDP data in Fig.~\ref{fig:HARPtotal}.  The ratio of
the Geant4 model prediction to the HARP or HARP-CDP result determines
how well we predict the data.

\subsection{Low-energy pion-production data}
\label{sec:abaev}
\begin{figure}
  \includegraphics[width = \columnwidth]{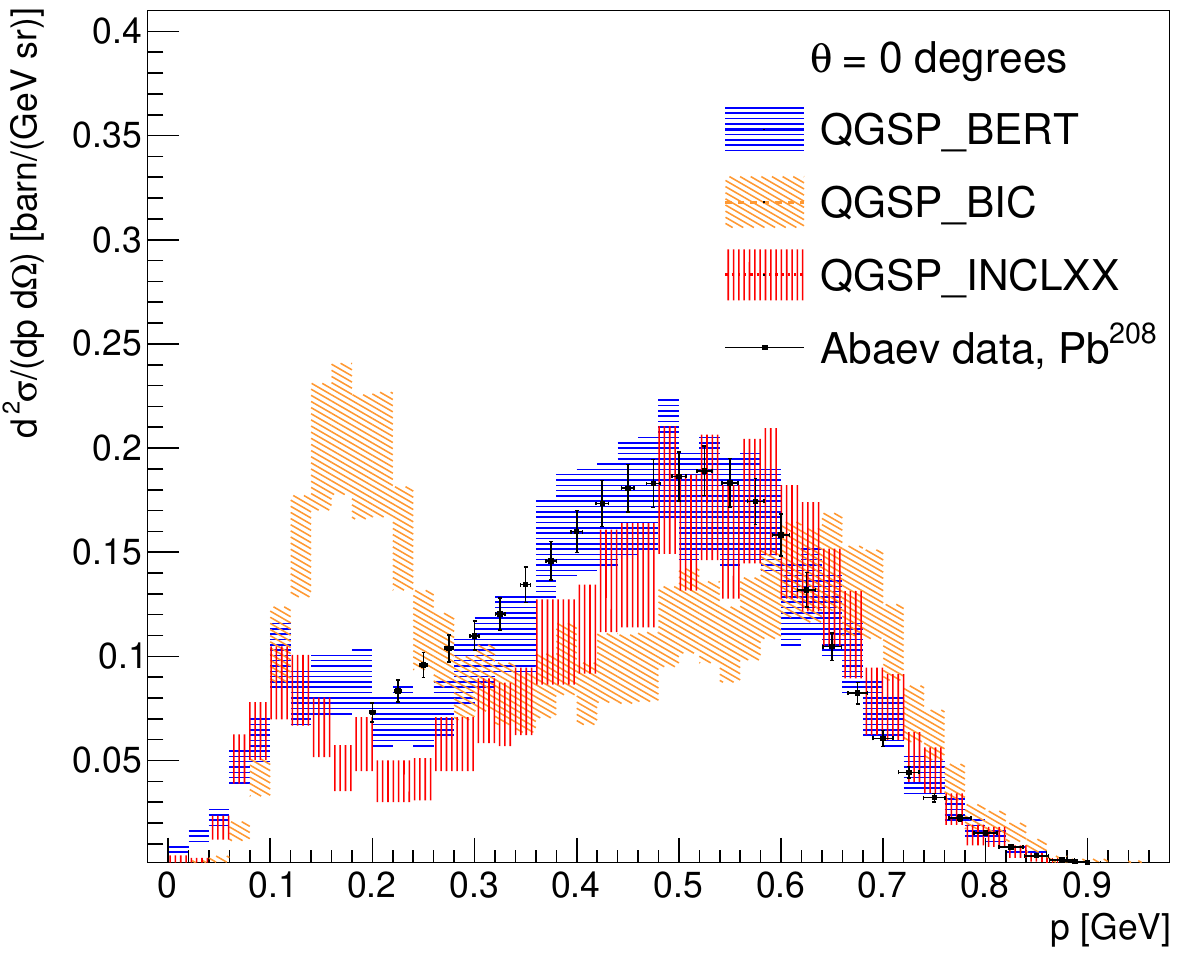}
  \includegraphics[width = \columnwidth]{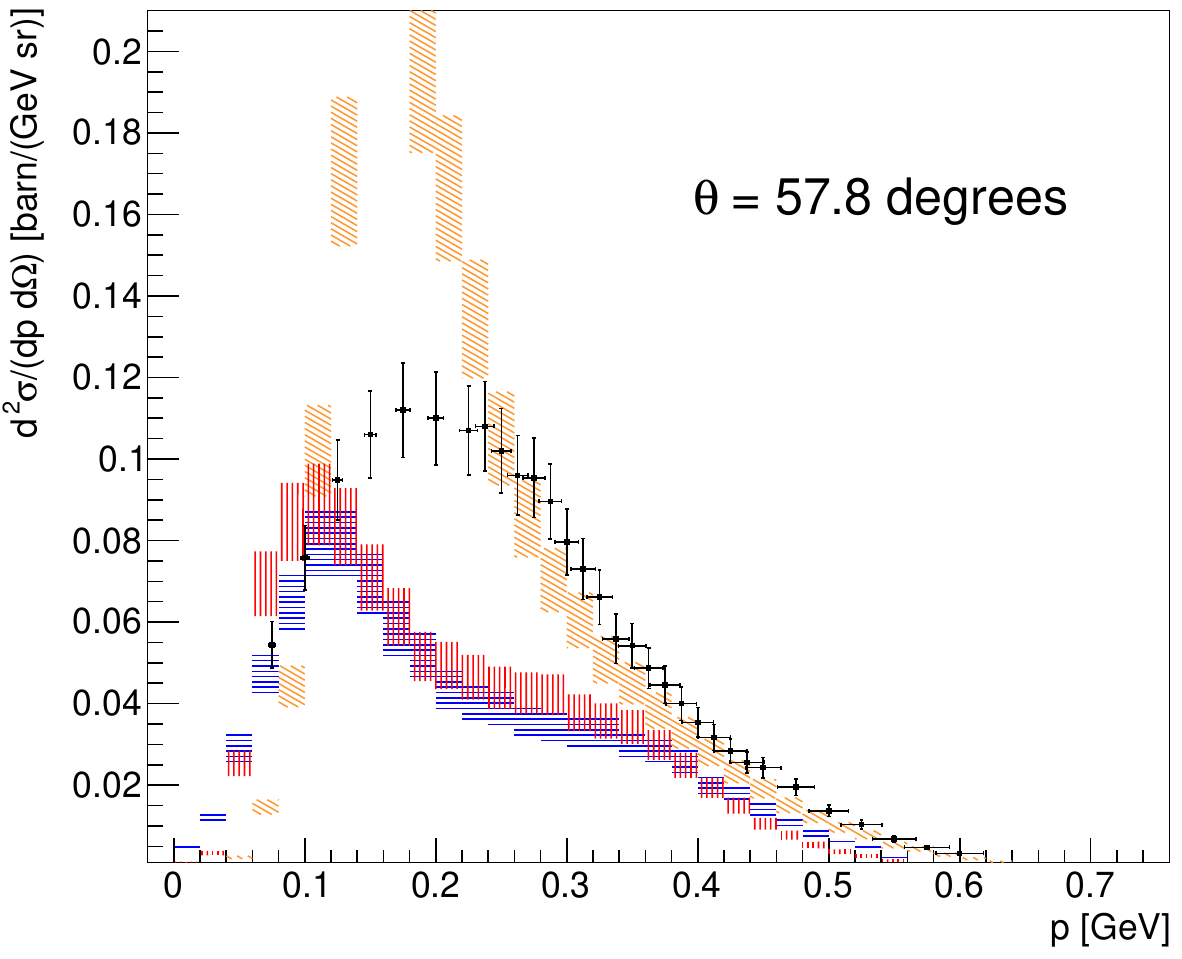}
  \caption{Comparisons of double-differential cross sections of
    $\pi^+$ production from 1~GeV p$ + ^{208}$Pb at 0$^{\circ}$ and
    57.8$^{\circ}$ as predicted by the different Geant4 physics lists
    to the measurements from Abaev \textit{et al.}}
  \label{fig:Abaev}
\end{figure}

\begin{figure}
  \includegraphics[width = \columnwidth]{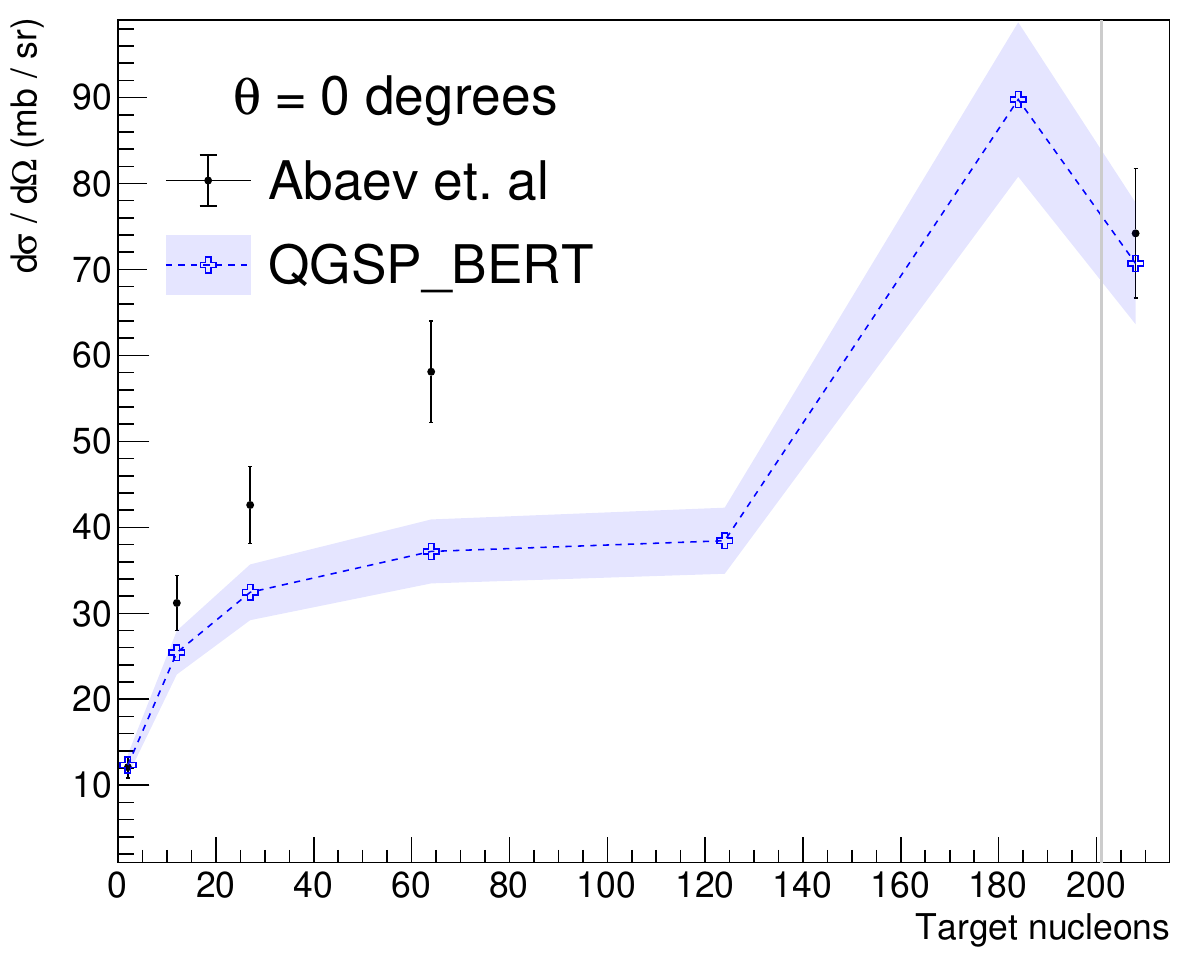}
  \includegraphics[width = \columnwidth]{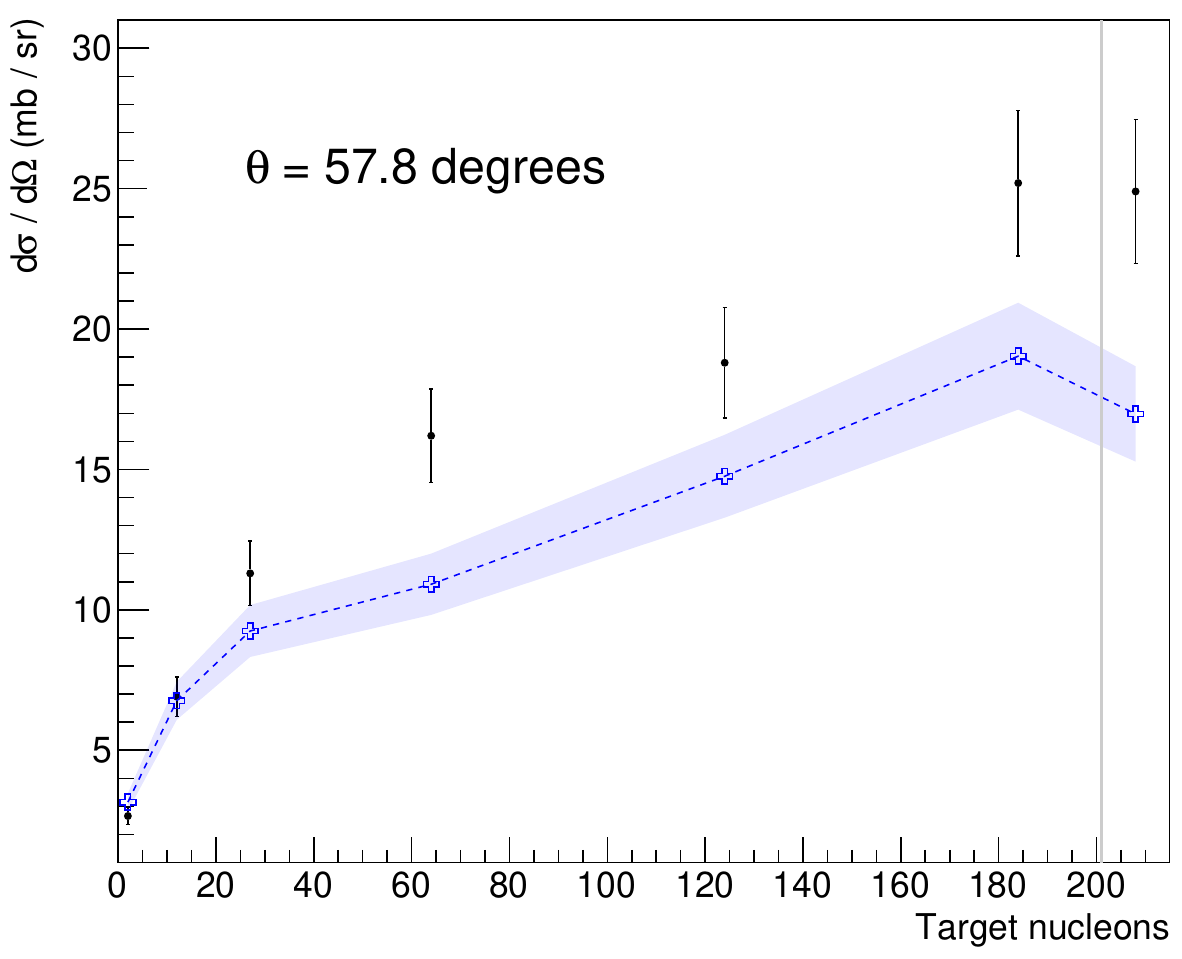}
  \caption{Comparisons of differential cross sections at 0$^{\circ}$
    and 57.8$^{\circ}$ as predicted by QGSP\_\hspace{2pt}BERT to
    measurements by Abaev \textit{et al.}  The vertical gray lines
    represent a mercury target.}
  \label{fig:AbaevAngle}
\end{figure}

Using the proton synchrotron at the Leningrad Nuclear Physics
Institute (Gatchina, Russia) with a beam kinetic energy of 997 $\pm$ 5
MeV, Abaev \textit{et al.}~measured pion production on 16 different
isotope targets at 0$^\circ$ and 57.8$^\circ$ with 0.01 steradian
angular acceptance~\cite{Abaev:1988az}.  We compare to a range of
nucleon numbers, but we exclude comparisons to different isotopes of
the same nucleus in this work as no significant difference between
different isotopes was found in the cross sections from data or
simulation.  The double-differential comparisons of the Geant4 models
to Abaev \textit{et al.}~are shown in Fig.~\ref{fig:Abaev}, and the
momentum-integrated comparisons are shown in Fig.~\ref{fig:AbaevAngle}.

\subsection{Secondary particle interactions}
\label{sec:thick}
The pion-production model of QGSP\_\hspace{2pt}BERT has the best
agreement for thin-target data, but we must also model the proton
energy loss and the interactions of any secondary particles that are
produced.  For example, $\pi^+$ scattering will affect our predictions
on how many $\pi^+$ decay at rest, and pion interactions such as
charge exchange or absorption will impact the number of $\pi^+$ that
decay into neutrinos.  In our simulations of the SNS, $\sim$~25\% of
the $\pi^+$ tracks that are produced end in non-decay processes
(labeled ``pi+Inelastic'' in Geant4).  Pions and other secondary
hadrons created at the SNS are well below 10~GeV and use the default
cross-section tables implemented in the Bertini Cascade model
(primarily the Baranshenkov and Glauber-Gribov parameterizations)
\cite{bertini}.  We do not perform any specific validation of these
processes in this work.

\subsection{Interpretation}
\label{sec:sys}

We are not aware of any data from p$ + $Hg and very few data sets
exist at these energies, so this work is intended as a cross-check of
prior estimates rather than as a derivation of our neutrino-flux
systematic.  We choose to simulate the SNS using
QGSP\_\hspace{2pt}BERT, and find that a 10\% uncertainty is consistent
with our validation studies.  In particular, QGSP\_\hspace{2pt}BERT is
the only model which agrees at the 10\% level with the cross section
measurements of both HARP and HARP-CDP; the other lists overpredict
the HARP-CDP data.  The validation against the Norbury-Townsend
parameterization further demonstrates an overall normalization problem
with QGSP\_\hspace{2pt}BIC, despite noteworthy agreement in the tails
of the 57.8$^{\circ}$ Abaev measurement.  While
QGSP\_\hspace{2pt}INCLXX is acceptable, QGSP\_\hspace{2pt}BERT has
better agreement with the data and the added bonus of being more
computationally efficient.  We note that while the momentum-integrated
Abaev data may disagree with QGSP\_\hspace{2pt}BERT predictions at
more than the 10\% level, similar disagreement is shown in the bottom
panel of Fig.~\ref{fig:singleDiffs} for single points of the
momentum-integrated HARP and HARP-CDP comparisons; it is only after an
additional integration over angle that good agreement is achieved.
Ultimately, the limited angular coverage of the Abaev data limits our
ability to investigate this effect.

In light of these studies and prior work using the Bertini cascade for
neutrino flux calculations \cite{lahetLSND, lahetKARMEN, lahetKARMEN2,
  numi, dune}, we continue to use QGSP\_\hspace{2pt}BERT with a 10\%
uncertainty on the flux predictions that come from our Geant4
simulations.  This systematic cannot be improved without new
measurements, and we describe future avenues for reducing this
uncertainty in Section \ref{sec:future}.

\section{Modeling the Spallation Neutron Source in Geant4}
\label{sec:geant4}

\begin{figure}
  \includegraphics[width = 0.85\columnwidth]{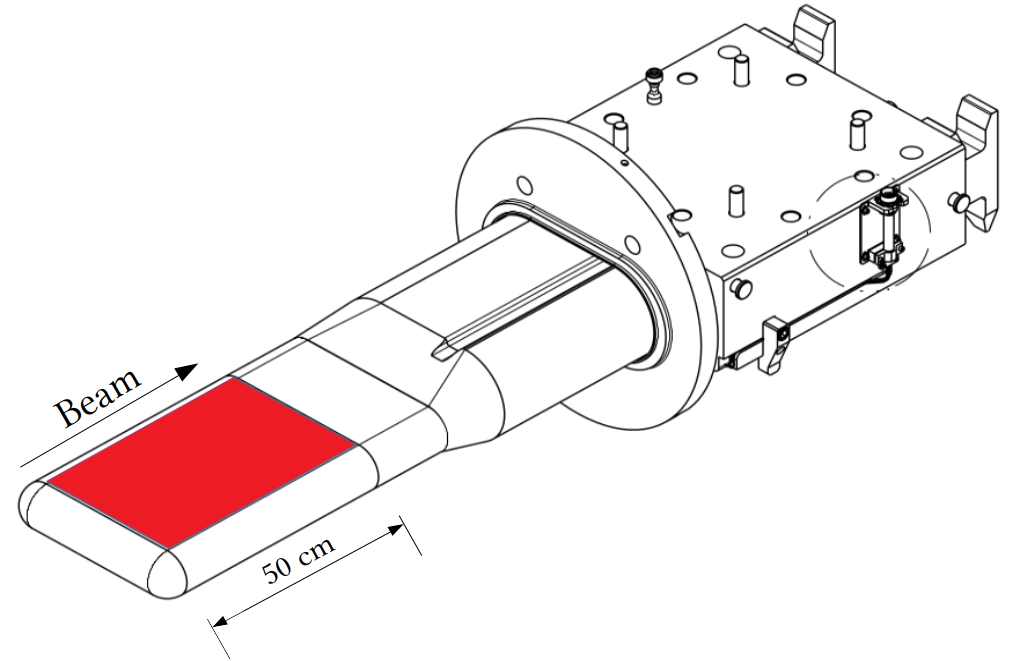} \quad
  \includegraphics[width = 0.95\columnwidth]{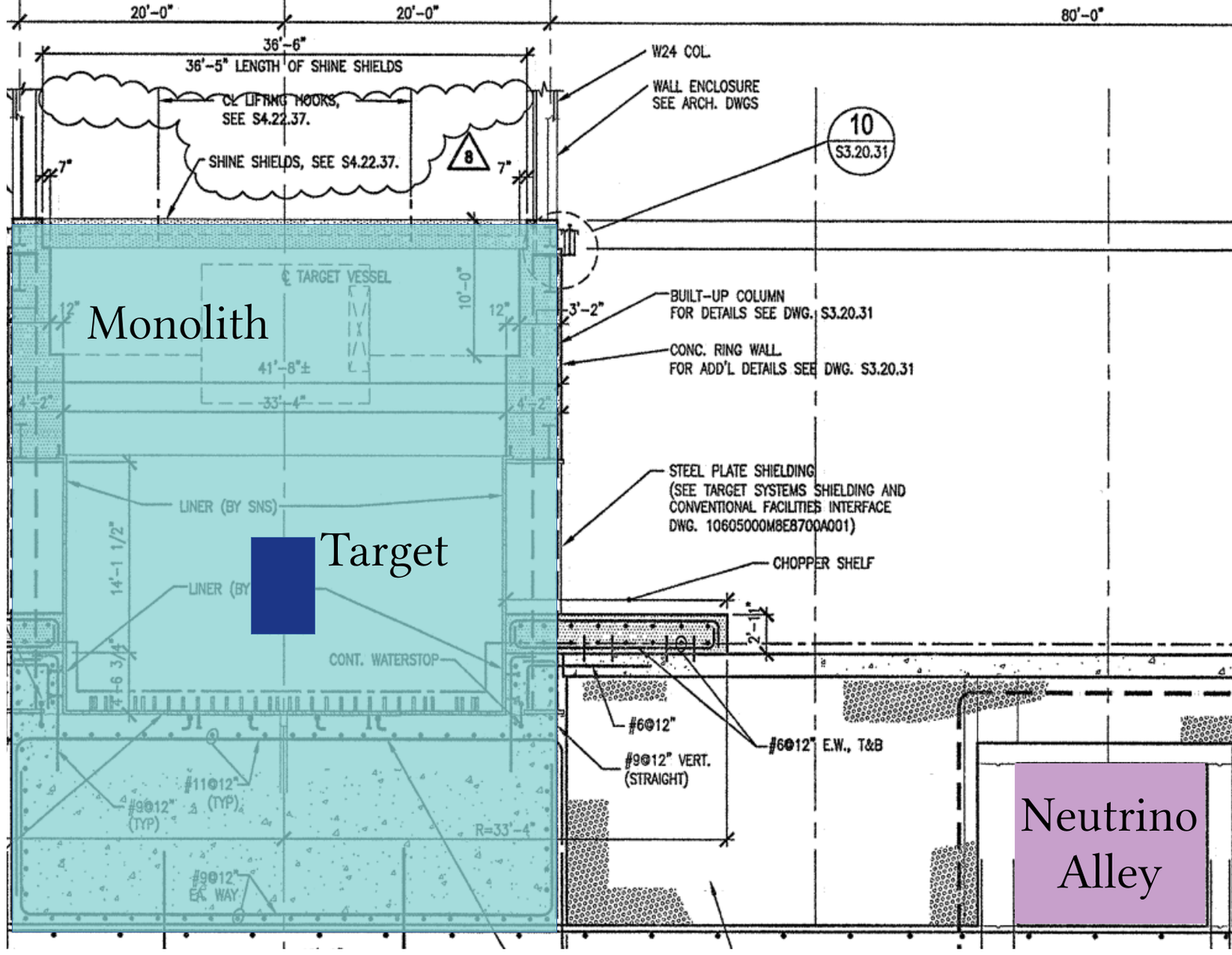}
  \caption{Top: ORNL technical drawing of the target vessel.  The red
    section highlights the main Hg target as implemented in our Geant4
    model.  Bottom: A portion of an ORNL technical drawing illustrating
    the target hall, with pieces in our Geant4 model highlighted.  The
    outer shaded cyan is the concrete monolith, with the inner indigo
    representing the steel containing the Hg target and moderators.
    In the bottom right corner, the shaded purple shows the location
    of Neutrino Alley relative to the target monolith
    \cite{ORNLdrawings}.}
  \label{fig:target}
\end{figure}

\begin{figure}
  \includegraphics[width = 0.85\columnwidth]{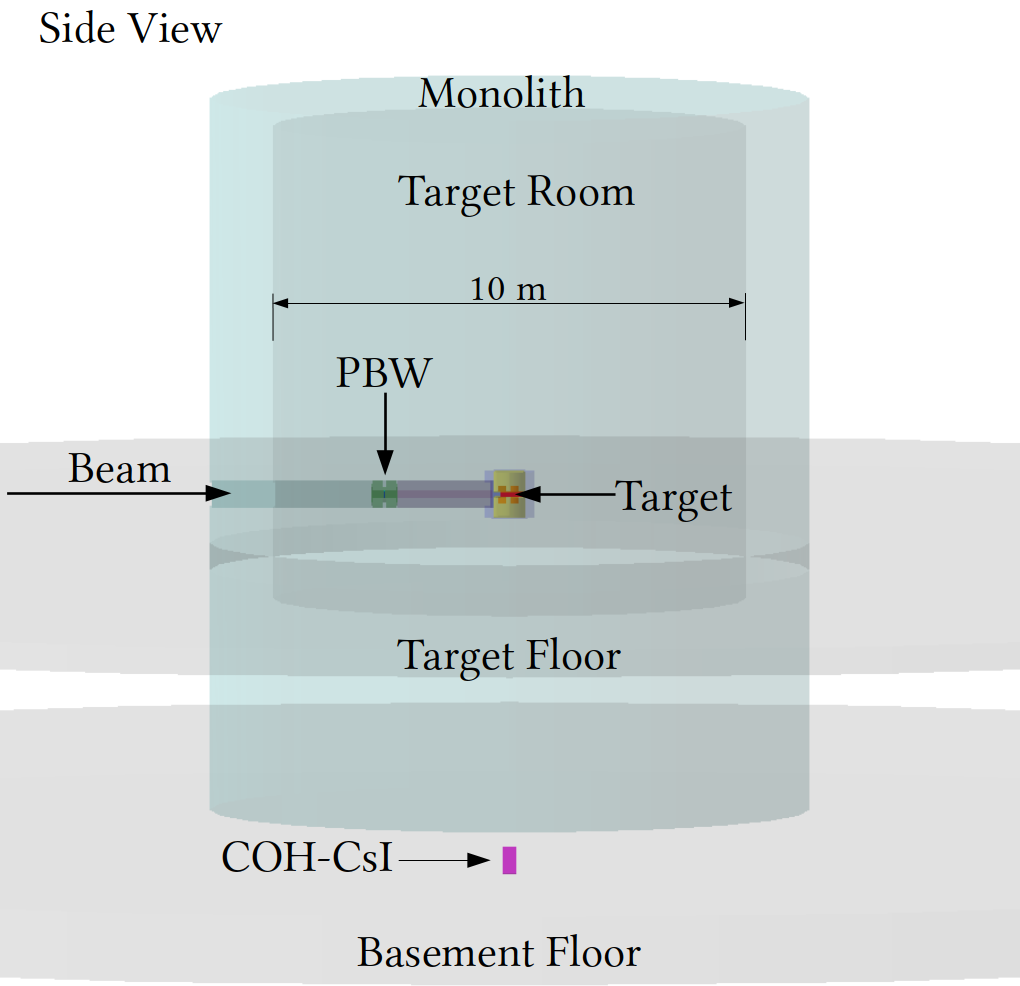}
  \includegraphics[width = 0.95\columnwidth]{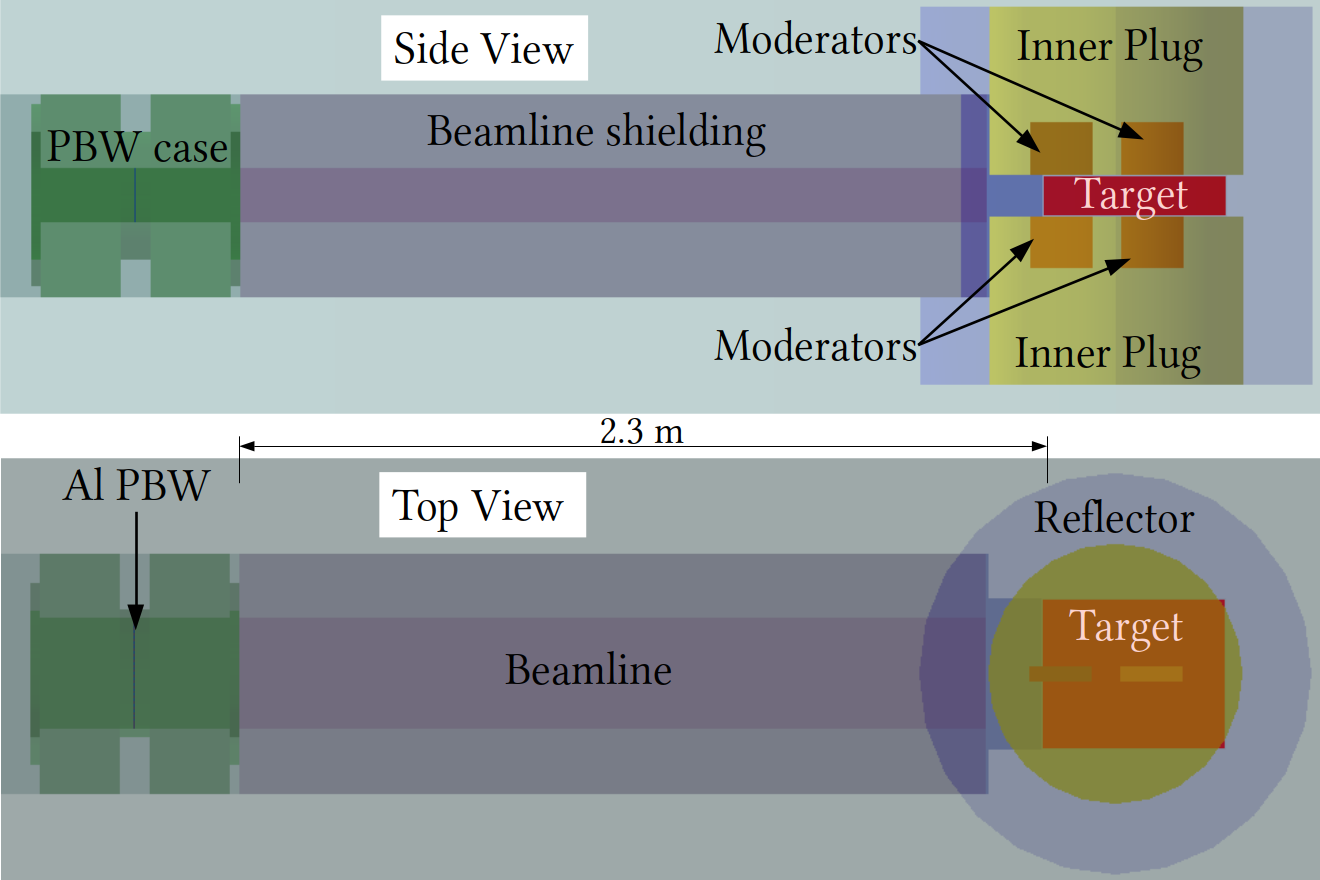}
  \caption{Our Geant4 model of the SNS, simplified from ORNL technical
    drawings.  Top: The full Geant4 world, highlighting the monolith
    relative to the location of COH-CsI in Neutrino Alley.  Bottom: A
    view inside the outer monolith illustrating the target, neutron
    moderator suite, proton beam window, and beamline shielding.}
  \label{fig:screenshots}
\end{figure}

The design of the SNS target and moderator suite was optimized for
neutron production and related science \cite{sns_design}.  We define
simplified components of the SNS target monolith that are expected to
contribute to pion production or to the stopping of pions and muons.
The simplification process is demonstrated in the top panel of
Fig.~\ref{fig:target}, where the technicalities of the target vessel
are reduced to the mercury-containing region shaded red.  The bottom
panel of Fig.~\ref{fig:target} highlights the target monolith and
Neutrino Alley to illustrate the structures we build into our model.
The details of our SNS model, along with their relative contributions
to the overall $\pi^+$ production, are shown in Table \ref{tab:dims},
and the full visualization of our simple model is shown in
Fig.~\ref{fig:screenshots}.

\begin{table*}
  \centering
  \caption{An overview of components in our Geant4 model that
    contribute to the overall pion-production.  We also include the
    fraction of $\pi^+$, and therefore $\nu$, our simulations produce
    as a result of each volume.  We report the dimensions from the
    perspective of the beamline as either Width $\times$ Height
    $\times$ Depth or Diameter ($\varnothing$) $\times$ Height.  The
    depth of the Inconel proton beam window (PBW) is an approximation
    (indicated by an *) of 3~cm, which includes some amount of vacuum
    immediately before and after the window as a result of the
    curvature.}
  \begin{tabular}{ccccc}\hline\hline
    \multirow{2}{*}{Component} & \multirow{2}{*}{Material} & \multirow{2}{*}{Dimensions} & \multicolumn{2}{c}{$\pi^+$ contributed}\\
    & & & \phantom{ss}Aluminum PBW & \phantom{ss}Inconel PBW\\\hline
    Target & Hg & 39.9 $\times$ 10.4 $\times$ 50.0~cm$^3$ & 94.12\% &\phantom{1}(90.91\%)\\ 
    Target Casing & Steel & 40.9 $\times$ 11.4 $\times$ 51.0~cm$^3$ & \phantom{9}0.20\% &\phantom{9} (0.56\%)\\ 
    Inner Plug (2) & Be, D$_2$O & 70.0~cm $\varnothing$, 45~cm & \phantom{9}0.19\% &\phantom{9} (0.23\%)\\ 
    Moderator (4) & H$_2$ (3), H$_2$O (1) & 4.0 $\times$ 13.9 $\times$ 17.1~cm$^3$ & \phantom{9}0.01\% &\phantom{9} (0.01\%)\\ 
    Reflector & Steel, D$_2$O & 108~cm $\varnothing$, 101.6~cm & \phantom{9}0.99\% &\phantom{9} (1.34\%)\\ 
    Beamline Shielding & Steel & 64.8 $\times$ 54.6 $\times$ 200.0~cm$^3$ & \phantom{9}0.93\% &\phantom{9} (1.72\%)\\ 
    Target Room & Steel & 1002~cm $\varnothing$, 950.8~cm & \phantom{9}0.00\% &\phantom{9} (0.14\%)\\ 
    Aluminum PBW & Al-6061, H$_2$O & 29.8 $\times$ 14.6 $\times$ 0.02~cm$^3$ & \phantom{9}2.77\% &\phantom{9} (--------)\\ 
    Aluminum Beamline & Air & 29.8 $\times$ 14.6 $\times$ 200.0~cm$^3$ & \phantom{9}0.79\% &\phantom{9} (--------)\\ 
    Inconel PBW & Inconel-718, H$_2$O& 26.7 $\times$ 12.7 $\times$ 3.0~cm$^3$* & \phantom{9}-------- &\phantom{9} (4.32\%)\\ 
    Inconel Beamline & Air & 26.7 $\times$ 12.7 $\times$ 200.0~cm$^3$ & \phantom{9}-------- &\phantom{9} (0.77\%)\\\hline\hline
  \end{tabular}
  \label{tab:dims}
\end{table*}

\par
\begin{figure}
\includegraphics[width = 0.49\columnwidth]{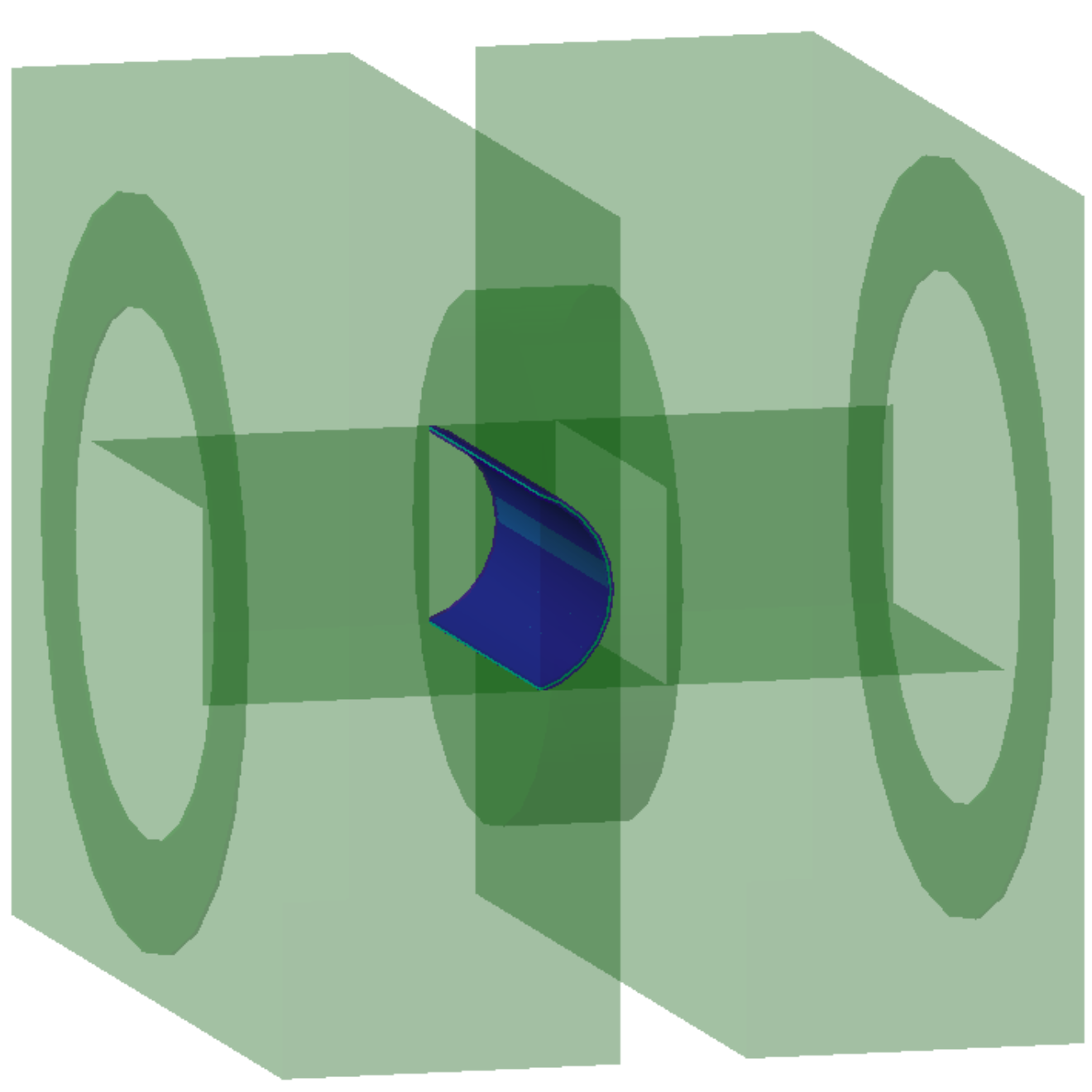}
\includegraphics[width = 0.49\columnwidth]{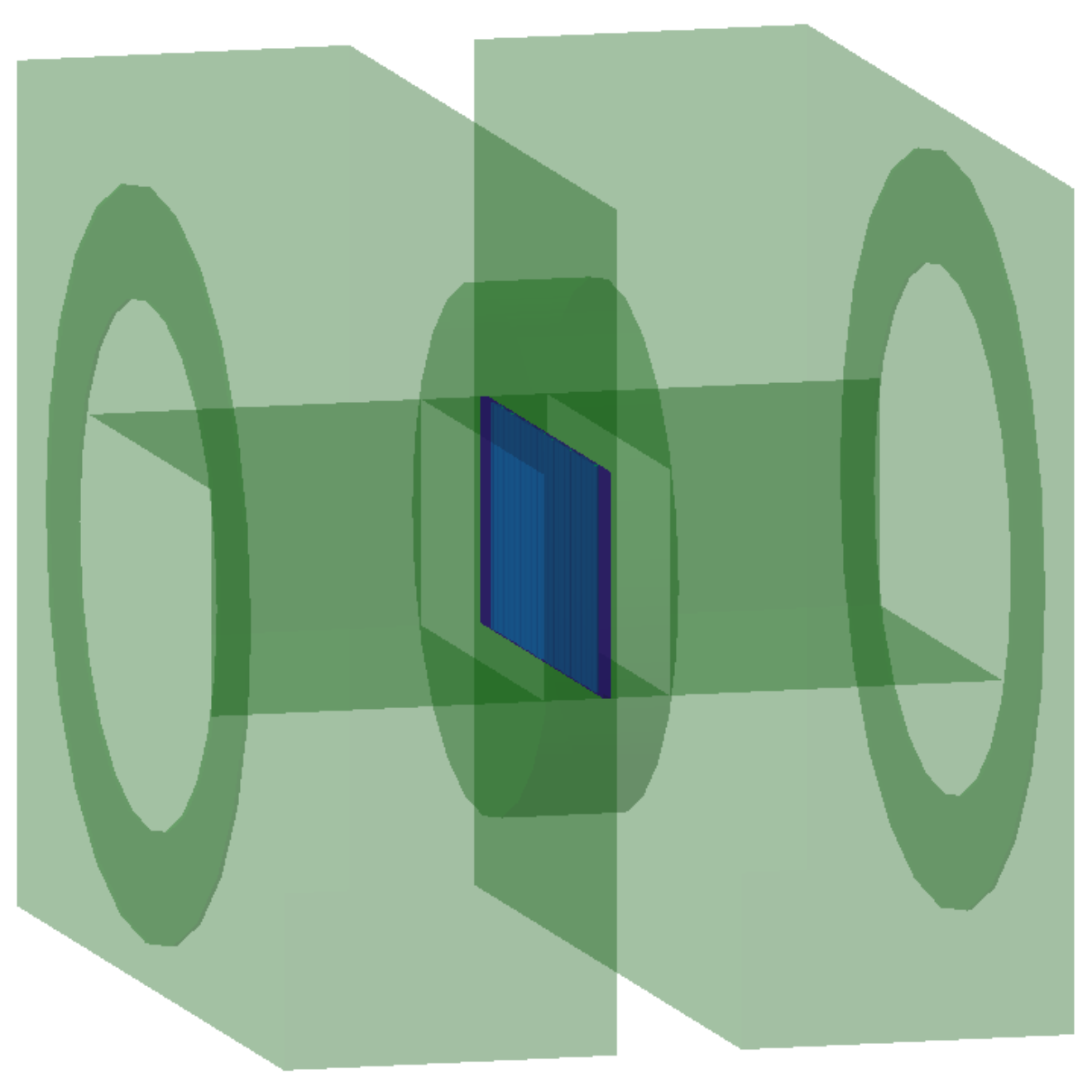}
\quad
\includegraphics[width = \columnwidth]{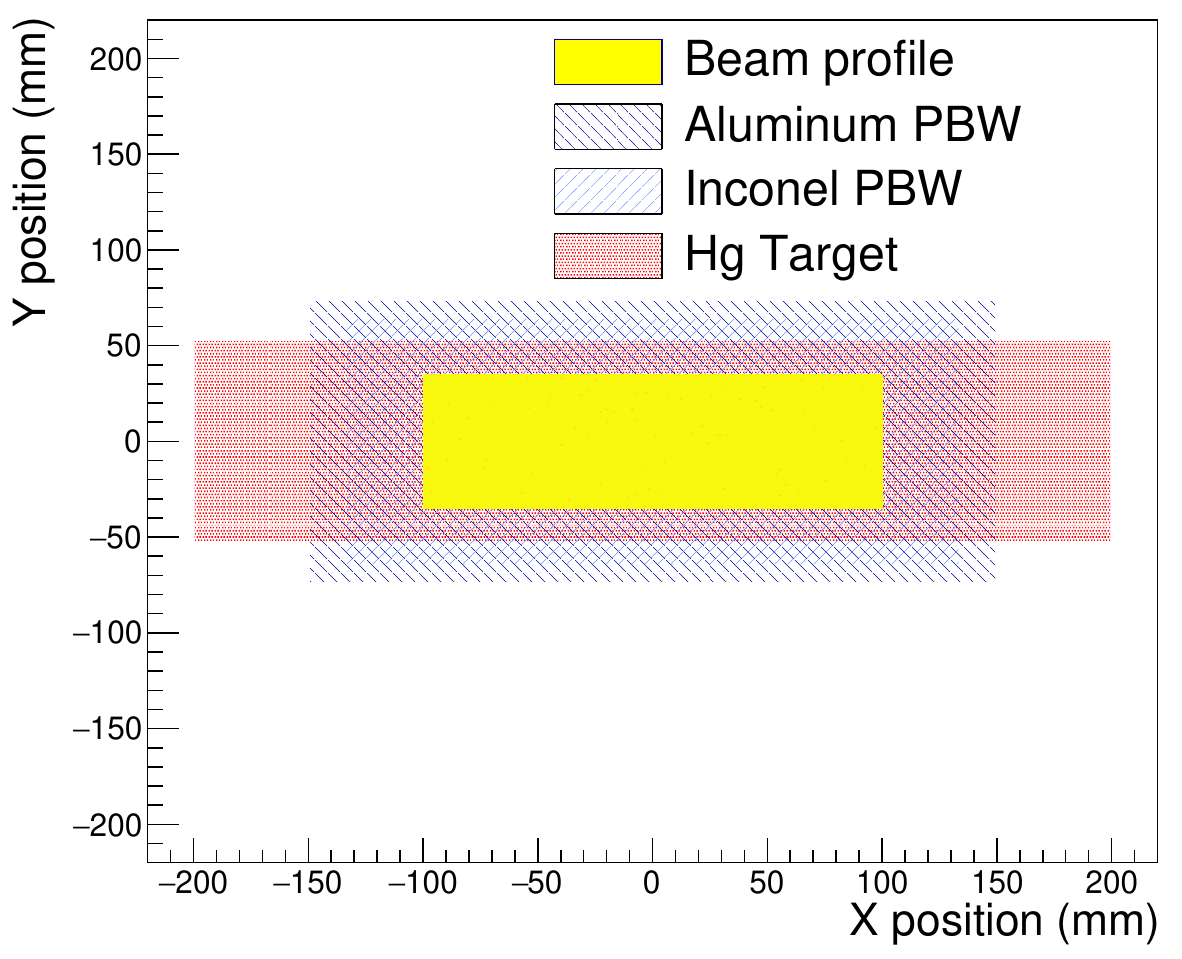}
\caption{Top left: Geant4 mockup of the dual-film Inconel PBW, with
  water cooling between the two films. Top right: Geant4 mockup of the
  aluminum plate PBW, with 50 vertical pipes for water cooling. Bottom:
  The position of incident protons shown relative to the profiles of
  the different PBW designs and Hg target.}
\label{fig:pbwAndProfile}
\end{figure}

\indent Though most of the components we simulate are essentially
unchanged during running despite routine maintenance and possible
replacements, we must carefully consider the proton beam window (PBW)
separating the vacuum of the accelerator from the target.  Each proton
must pass through the PBW, resulting in both proton energy loss and
pion production as a result of interactions in the thin window.  The
PBW is routinely replaced due to radiation damage, and two different
PBW designs have been in use during COHERENT's live-time in Neutrino
Alley.  A two-layered film design using Inconel, a nickel-based alloy
trademarked by the Special Metals Corporation~\cite{inconel}, with
water cooling between the films, was used from the initial SNS
production runs until January 11, 2017.  An aluminum plate design with
50 drilled pipes for water cooling was in place until the latest
replacement reverted back to an Inconel PBW on April 7, 2020.
Figure~\ref{fig:pbwAndProfile} illustrates both PBW designs as modeled
in our Geant4 geometry.

\par\indent The SNS accelerates protons into an accumulator ring,
which ensures that a focused beam of monoenergetic protons is directed
on to the target.  This beam is magnetically spread to prevent
overheating of the proton beam window and target casing
\cite{beamSpread}; we introduce a uniformly distributed source using a
prior measurement of the beam profile at the target
\cite{blokland2010sns} to account for this.  Our simulated profile is
illustrated in the bottom panel of Fig.~\ref{fig:pbwAndProfile} to
show its size relative to the beam window designs and target.

\par\indent We specify the particles for the simulation to track,
typically $\nu_x$, $\pi^\pm$, $\mu^\pm$, $K^\pm$, $\eta$, $p$, and
$n$, to ensure that we do not truncate any possible neutrino
production chain.  Using the Monte Carlo framework of Geant4 and the
QGSP\_\hspace{2pt}BERT physics model chosen in Section
\ref{sec:physlists}, we observe which particles and interactions are
responsible for generating the SNS neutrino flux.  The predictions we
make are dependent on our chosen physics model; for example, the
QGSP\_\hspace{2pt}BIC nuclear model predicts some production of the
$\eta$ meson given 1~GeV incident protons while other models do not.

\section{Simulated neutrino flux for the First Target Station}
\label{sec:fts}

\begin{figure}
  \includegraphics[width = \columnwidth]{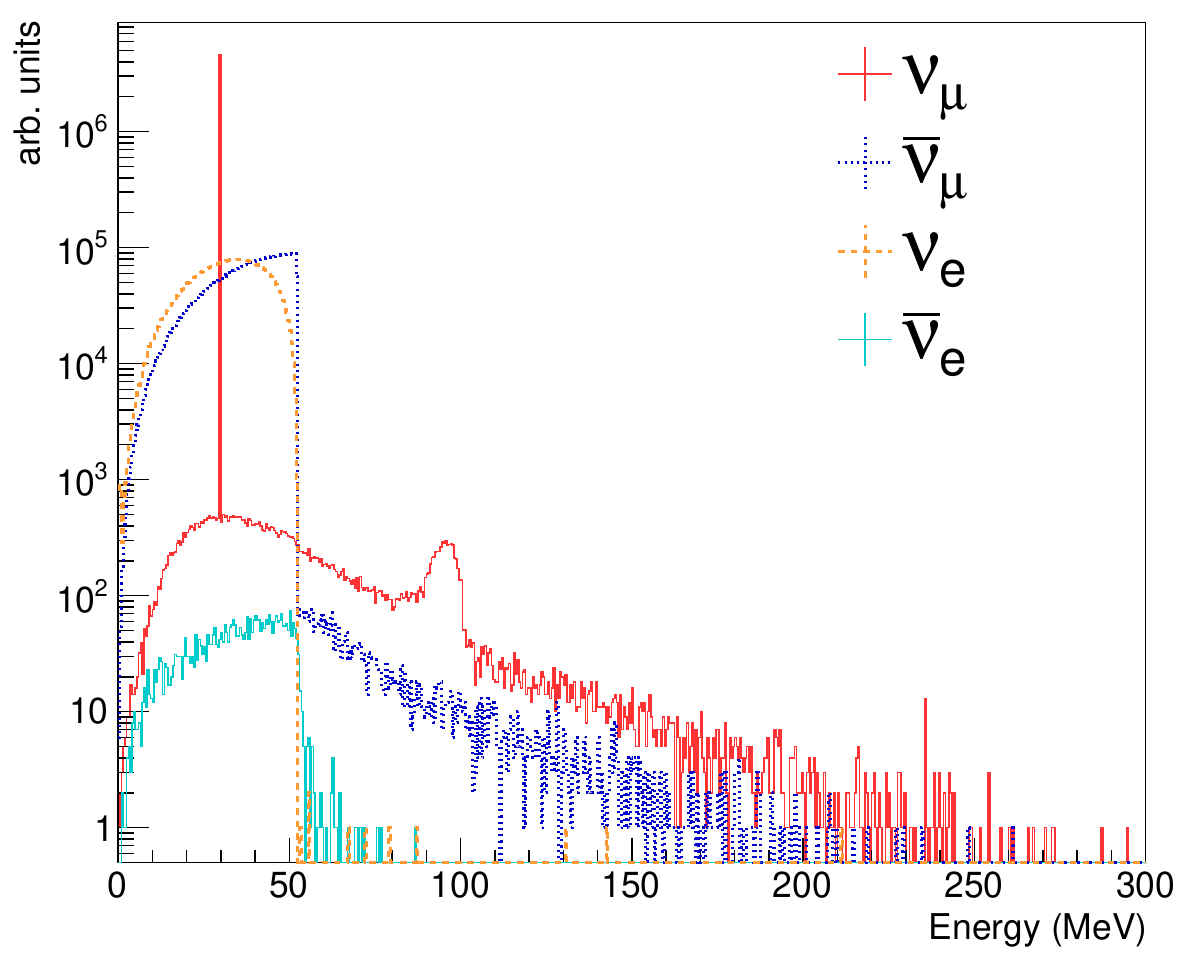}
  \includegraphics[width = \columnwidth]{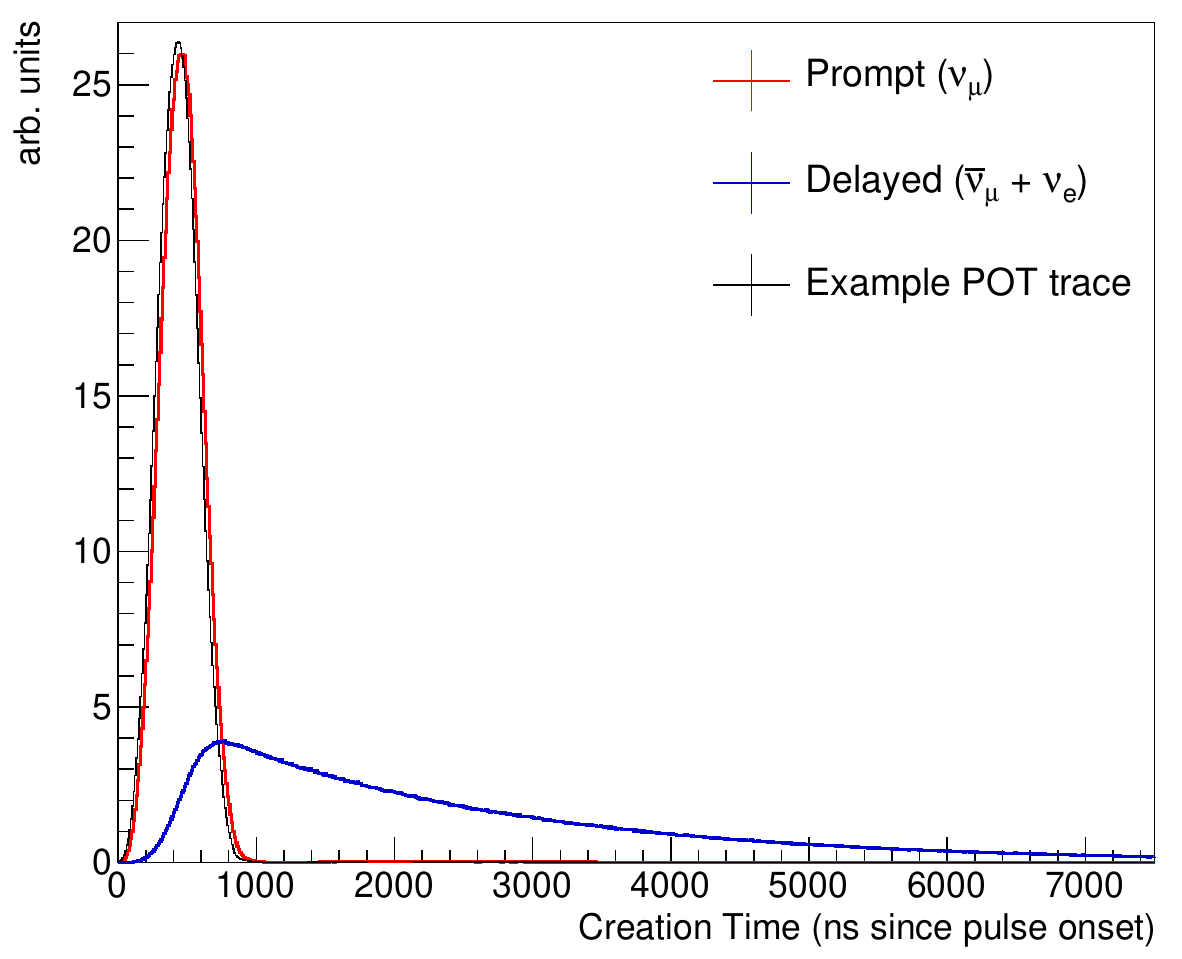}
  \caption{Distributions of neutrino energy (top) and creation time
    (bottom) produced at the SNS, using QGSP\_\hspace{2pt}BERT to
    model the interactions of 1~GeV protons incident on the aluminum
    PBW geometry.  We convolve the single proton output of our
    simulations with the proton-on-target trace.}
  \label{fig:spectra}
\end{figure}

\begin{figure}
  \includegraphics[width = 0.95\columnwidth]{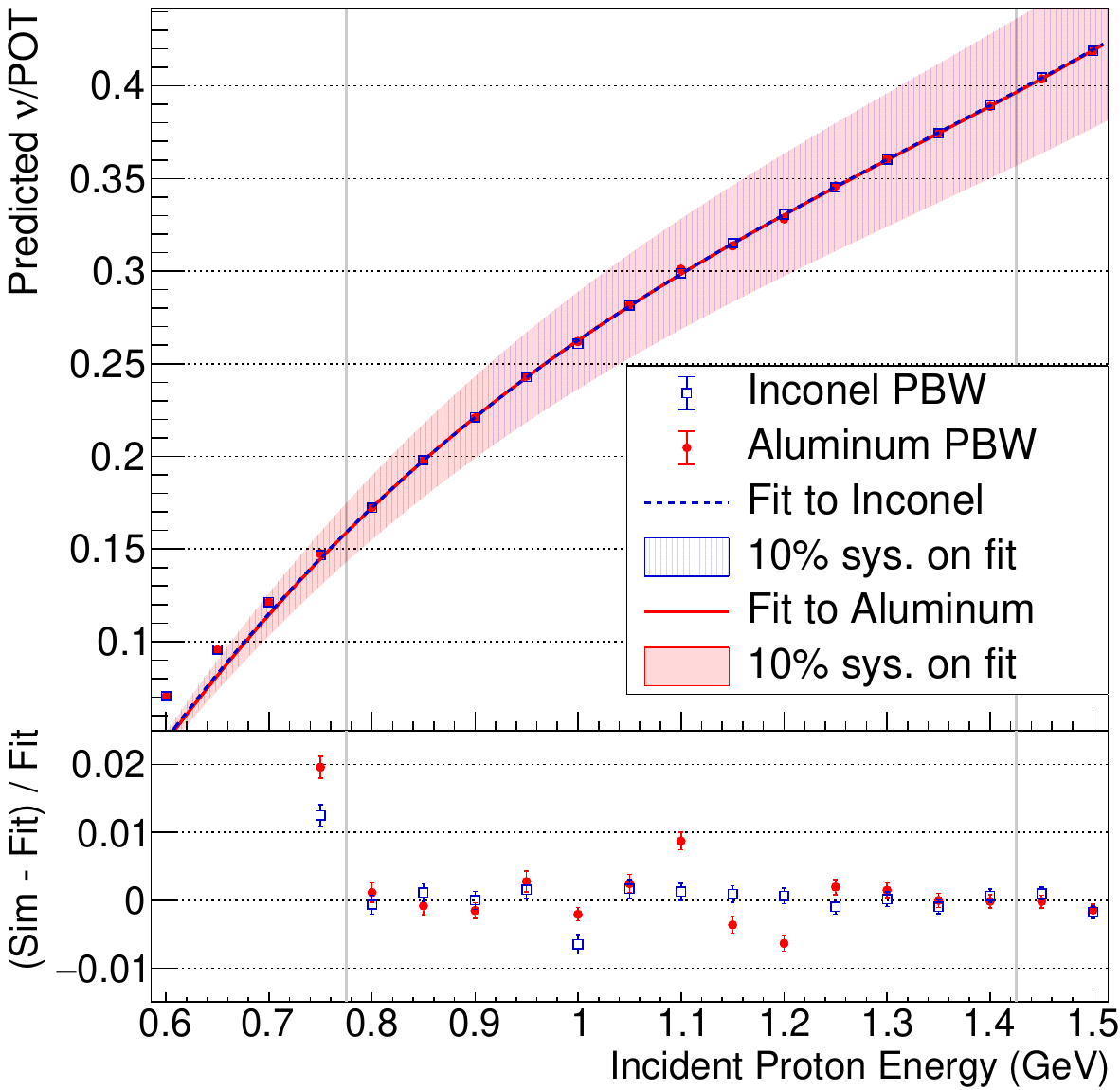}
  \includegraphics[width = 0.95\columnwidth]{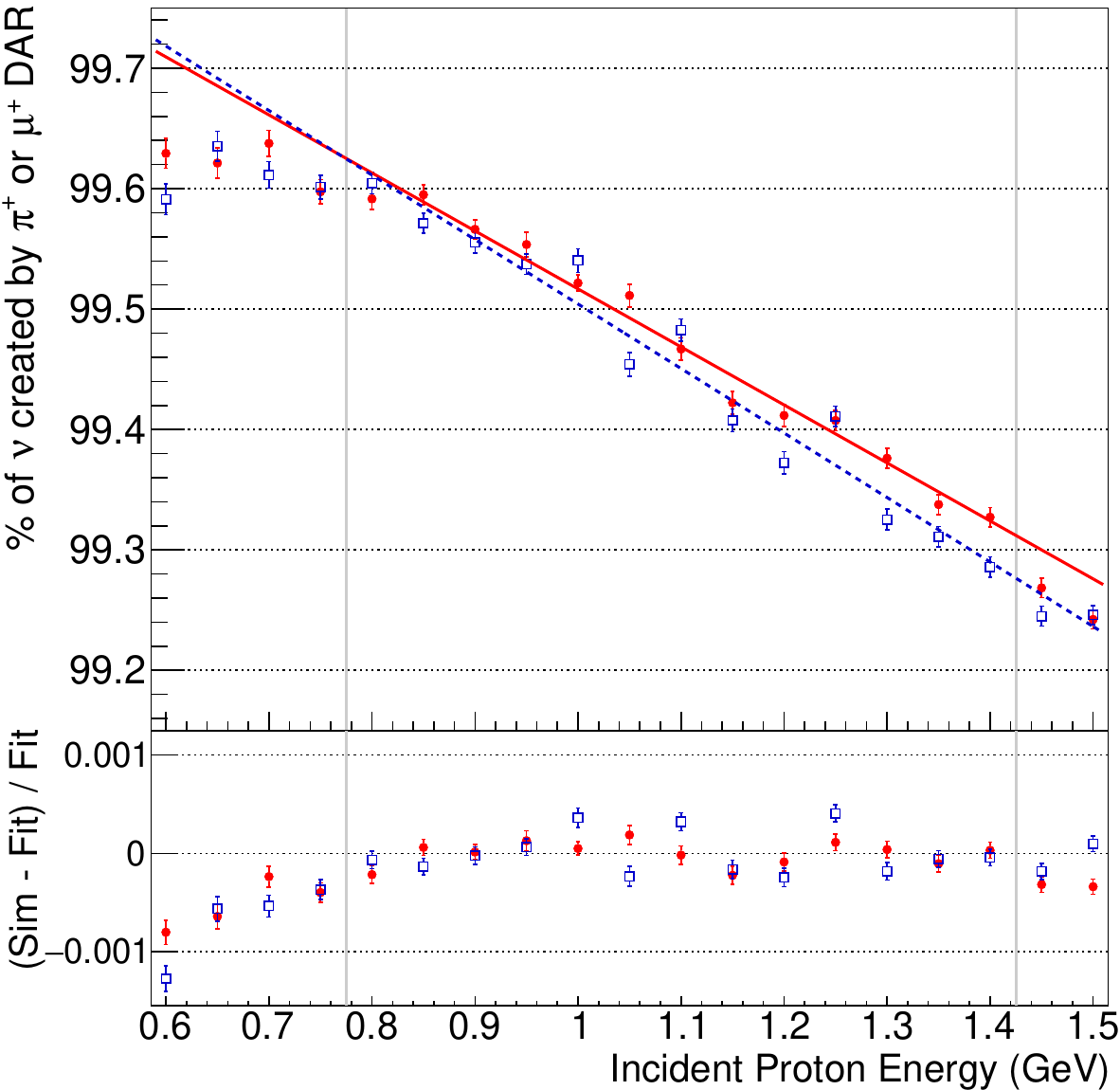}
  \caption{Top: The total neutrino flux from the SNS will depend on
    the incident proton energy, and each operational configuration
    demonstrates a cubic dependence on this parameter.  Bottom: The
    fraction of neutrinos produced from decay-at-rest processes
    demonstrates a linear dependence on the incident proton energy
    above $\sim$~0.8~GeV.  The fit range for both plots is $E \in
    [0.775, 1.425]$~GeV; this is the region between the vertical gray
    lines.  The bottom panel in each plot shows the relative
    residuals, calculated as in the axis label from the simulation
    (``Sim'') and fit (``Fit'') predictions.}
  \label{fig:protonEnergyDep}
\end{figure}

\begin{table*}
  \centering
  \caption{A breakdown of the processes and parent particles which
    create neutrinos for 1~GeV protons at the SNS with an aluminum
    PBW.  The creation processes are classified as decay at rest
    (DAR), decay in flight (DIF), $\mu^-$ capture, or decay in orbit
    (DIO).  We include significant figures here to sum to 100\% given
    the small contributions outside of the $\pi^+$ DAR chain.}
  \label{tab:breakdown}
  \begin{tabular}{ccccccccccccc}\hline\hline
    & &\multirow{2}{*}{$\nu$ / POT}  & & \multicolumn{4}{c}{Creation Process} & & \multicolumn{3}{c}{Parent Particle}\\
    & \phantom{ss} & & \phantom{ss} & DAR & DIF & $\mu^-$ Cap & $\mu^-$ DIO & \phantom{ss} & $\pi^+$ or $\mu^+$ & $\pi^-$ or $\mu^-$ & $K^+$ \\\hline
    $\nu_\mu$ & & 0.0875 & & 98.940\% & 0.779\% & 0.196\% & 0.084\% & & 99.7185\% & 0.2812\% & 0.0003\% \\
    $\bar{\nu}_\mu$ & &0.0875 & & 99.718\% & 0.282\% & --- & --- & & 99.7187\% & 0.2813\% & --- \\
    $\nu_e$ & &0.0872 & & 99.999\% & 0.001\% & --- & --- & & 99.9999\% & --- & 0.0001\% \\
    $\bar{\nu}_e$ & &0.0001 & & --- & 0.331\% & --- & 99.669\% & & --- & 100\% & --- \\\hline\hline
  \end{tabular}
\end{table*}

Figure \ref{fig:spectra} shows the energy and timing spectra for each
neutrino flavor present in the simulation using the
QGSP\_\hspace{2pt}BERT physics list to simulate incident protons with
1~GeV of kinetic energy on the SNS geometry with an aluminum PBW.  We
find that the SNS $\nu$ flux predictably demonstrates the
characteristics of a pion decay-at-rest source such as the
monoenergetic $\nu_\mu$ at $\sim$~30~MeV from $\pi^+$ decay at rest
and $\bar{\nu}_\mu$ and $\nu_e$ following the Michel spectra predicted
from the three-body $\mu^+$ decay at rest (DAR).  Variations from
these spectra include decays in flight (DIF), decays in orbit (DIO),
and $\mu^-$ capture.  We also observe some contribution from
decay-at-rest kaons, notably in the $\nu_\mu$ spectrum at
$\sim$~240~MeV, but due to the small phase space available to produce
these more massive particles, kaons have an almost negligible
contribution to the SNS neutrino flux. Ultimately, this simulation
predicts a decay-at-rest neutrino source with greater than $99\%$
purity, with the exact creation process and parent particle breakdown
shown for the aluminum PBW in Table \ref{tab:breakdown}.

Using 1~GeV protons incident on our SNS geometry from behind the
PBW, our simulations predict 0.262 neutrinos per proton on target.  We
find that our model of the SNS neutrino flux is primarily comprised of
$\nu_\mu$, $\bar{\nu}_\mu$, and $\nu_e$ (each greater than 0.087
$\nu_X$/POT, where $X = \mu, \bar{\mu}, e$) with a small contribution
of $\bar{\nu}_e$ (0.0001 $\bar{\nu}_e$/POT, not considering the
activation of materials near the target).  We also see a small flux of
low-energy $\bar{\nu}_e$ from neutron $\beta$-decay that we neglect in
this work, with the intention of performing a dedicated study of
radioactive products produced as a result of SNS operations in the
future.

COHERENT deployed detectors at the SNS prior to the accelerator
systems reaching 1~GeV, so data taken at lower energies
($\sim$~850~MeV) must also be understood.  The upcoming Proton Power
Upgrade \cite{ppuDesign} will prepare the SNS for the planned Second
Target Station (described in Section~\ref{sec:sts}) by improving the
accelerator.  The upgrade will see the SNS operate at a more intense
2.0~MW, with 1.3~GeV incident protons by 2024.  We use this simulation
to study the dependence of the neutrinos produced on the incident
proton energy and to develop an approach to account for changes to SNS
operations over a run period.  Figure \ref{fig:protonEnergyDep} shows
the energy dependence for both the total neutrino production and the
fraction of neutrinos produced by the $\pi^+$ decay chain, and the
parameters for each of the fits are listed in Table \ref{tab:params}.
This figure also demonstrates that while there are minimal differences
in total neutrino production between the two PBW designs, the
differences in the relative contribution of pion production resulting
from interactions with the PBW (see Table \ref{tab:dims}) can impact
the stopping power of the SNS.  The neutrino luminosity from the SNS
given particular operating conditions can then be calculated as

\begin{equation}
  \frac{\nu}{t} = \frac{\nu}{\textrm{POT}}\frac{\textrm{POT}}{t} = \frac{\nu}{\textrm{POT}}\frac{E_{\textrm{total}}}{E}\frac{1}{t} = F(E) \frac{P}{E},
  \label{eq:luminosity}
\end{equation}

\begin{table*}
  \centering
  \caption{Fit parameters for the proton-energy dependence studies
    using both beam window designs.  The three parameters for the
    cubic fits used in Eqn.~\ref{eq:luminosity} ($F(E) = p_3 E^3 + p_2
    E^2 + p_1 E + p_0$) are illustrated in the top panel in
    Fig.~\ref{fig:protonEnergyDep}, while the two parameters for the
    linear fits ($m E + b$) are illustrated in the bottom panel.  The
    fit uncertainties do not consider the overall 10\% systematic.}
  \label{tab:params}
  \begin{tabular}{ccccccccc}\hline\hline
     Design & \phantom{ssss} & $p_3$ [GeV$^{-3}$] & $p_2$ [GeV$^{-2}$] & $p_1$ [GeV$^{-1}$] & $p_0$ & \phantom{ssss} & $b$ & $m$ [GeV$^{-1}$] \\\hline
    Aluminum PBW & & 0.28(2) & -1.12(6) & 1.79(6) & -0.68(2) & &  99.99(1) & -0.48(1)\\
    Inconel PBW & & 0.27(2) & -1.09(6) & 1.75(6) & -0.67(2) & & 100.04(1) & -0.53(1)\\\hline\hline
  \end{tabular}
\end{table*}

\par\noindent where $E$ is the kinetic energy per proton, $F(E)$ is
the fraction of $\nu$ produced per proton on target with incident
kinetic energy $E$, $E_{\textrm{total}}$ is the combined energy of all
protons incident on the target in time $t$, and $P$ is the SNS beam
power ($E_{\textrm{total}} / t$).  Figure \ref{fig:protonEnergyDep}
demonstrates that $F(E)$ can be estimated as a cubic polynomial in $E$
with parameters defined in Table~\ref{tab:params}, for $E$ between
0.775 and 1.425 GeV.  Plugging this into Eqn.~\ref{eq:luminosity},
we find a general expression for the SNS neutrino luminosity:

\begin{equation}
  \frac{\nu}{t} = P\left(p_3\hspace{2pt}E^2 + p_2\hspace{2pt}E + p_1 + \frac{p_0}{E}\right).
\end{equation}

Using this functional form and typical pre-upgrade operational
parameters of 1.4~MW (7.0~GWhr/yr) and incident protons with 1~GeV of
kinetic energy, we calculate $2.36\times10^{15}$ neutrinos produced
per second while the SNS is running.  Estimating this production as an
isotropic point source, we calculate a neutrino flux of $4.7 \times
10^7~\nu$~cm$^{-2}$~s$^{-1}$ at 20 meters from the target center (the
approximate location of the first CEvNS measurements in COH-CsI).
Using the nominal SNS running time of 5000 hours per year, the SNS
sees $1.58\times10^{23}$ POT per year, with a $\nu$ luminosity of
$4.25\times10^{22}~\nu$ per year, or a flux of $8.46 \times
10^{14}~\nu$~cm$^{-2}$~yr$^{-1}$ at 20~m from the target.

\begin{figure}
  \includegraphics[width = \columnwidth]{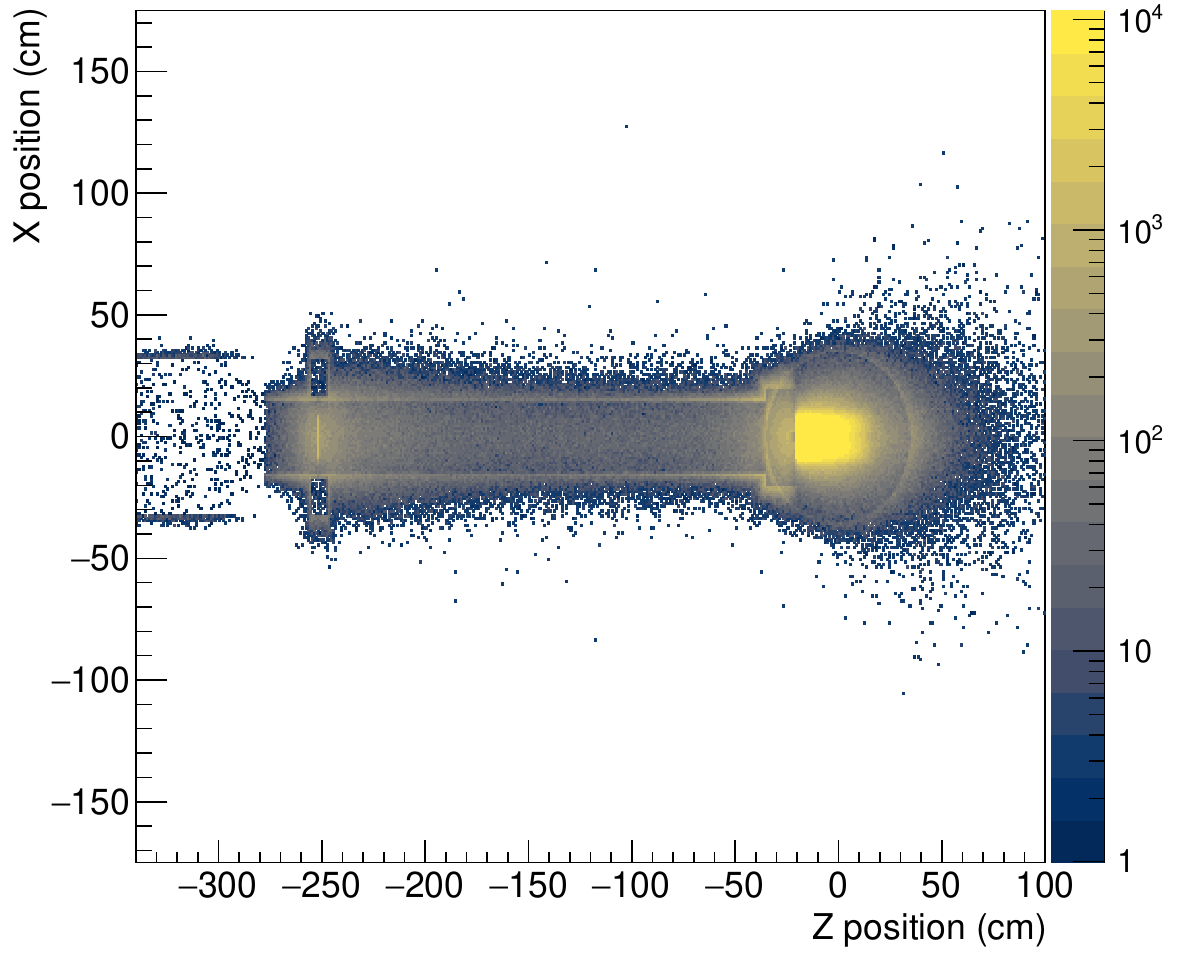}
  \includegraphics[width = \columnwidth]{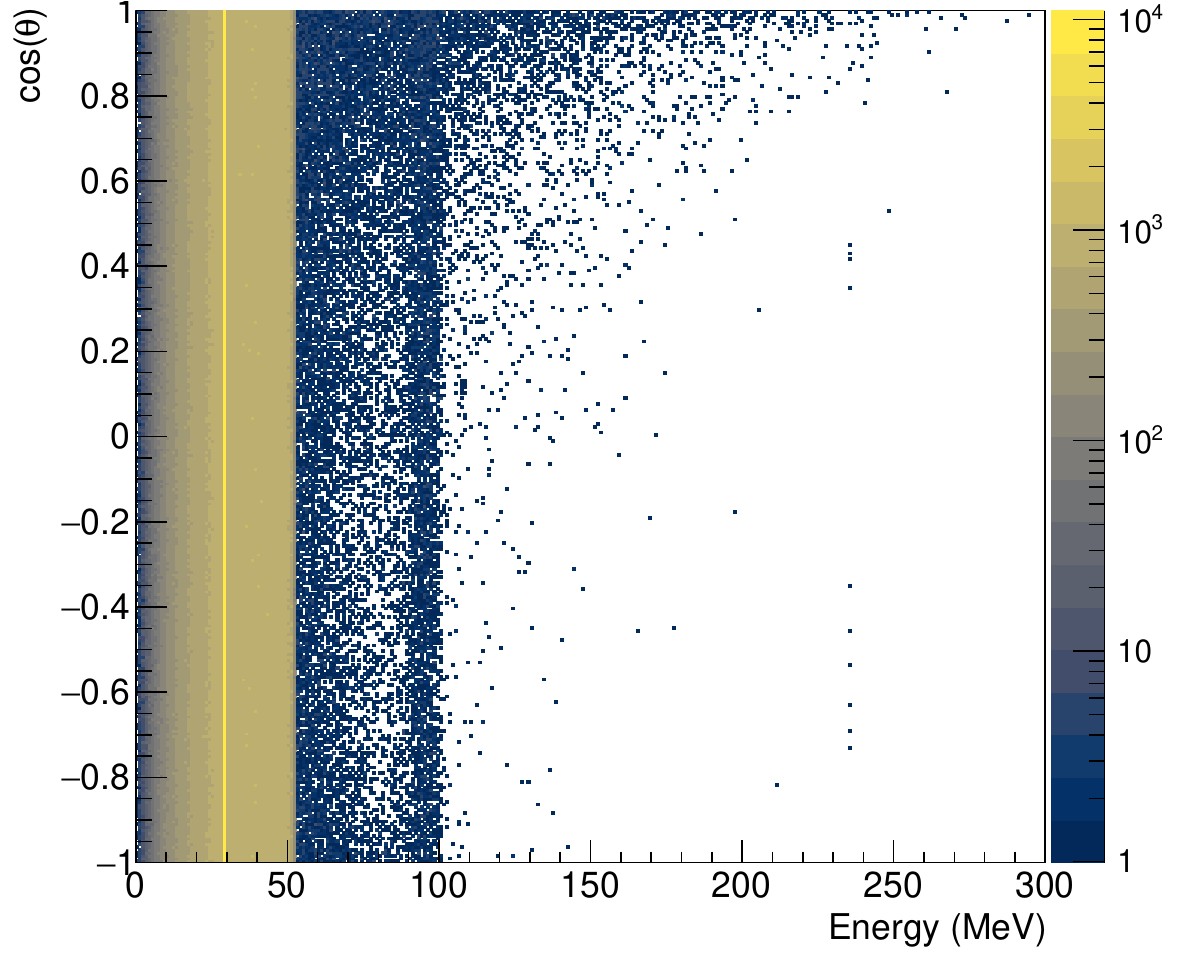}
  \caption{Top: A top-down view of the neutrino creation positions.
    Bottom: Distribution of the kinetic energies and production angles
    (relative to the beamline axis without convolving the creation
    position information from the top panel) of all neutrinos.}
  \label{fig:positions}
\end{figure}

We also study the creation positions and momenta of the neutrinos,
shown in Fig.~\ref{fig:positions}.  The volumes and materials which
create the pions were listed in Table \ref{tab:dims}; the neutrinos
are primarily produced after the short movements of pions and muons
coming to rest.  The spread of the beam and the movements of the
particles result in a radial spread from the beamline axis.  Over 86\%
of the neutrinos are produced within 10~cm of the beamline axis, and
almost all production ($>$99\%) occurs within 0.5~m of the beamline
axis.  Along the beamline axis, we find that over 90\% of the neutrino
production occurs within the target and less than 5\% of the neutrinos
are produced at the PBW location 2.5~m upstream of the target.
Because the $\pi^+$ and $\mu^+$ decay at rest, we also have almost
fully isotropic production of neutrinos up to about $50~$MeV.  We do
note visible anisotropy in the bottom panel of
Fig.~\ref{fig:positions} for $E_\nu~>~60~$MeV that is consistent
with neutrinos boosted in the forward direction from pions decaying in
flight.

\begin{figure*}
  \includegraphics[width = \textwidth]{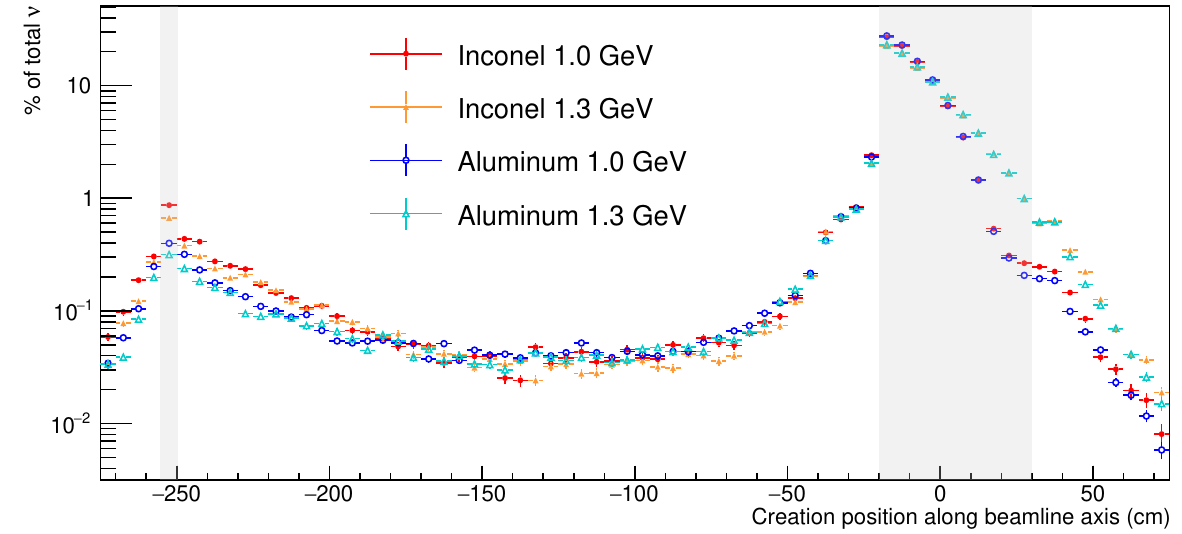}
  \caption{A comparison of neutrino production along the beamline for
    different SNS configurations and beam energies.  The gray shading
    to the left indicates the position of the proton beam window, and
    the shading to the right indicates the position of the Hg
    target.}
  \label{fig:cumulative}
\end{figure*}

\par We find that both PBW designs cause some neutrino production
outside of the target regardless of the incident proton energy as
illustrated in Fig.~\ref{fig:cumulative}.  However, our detectors are
deployed $\sim$~20~m from the target center, with the PBW placement
only 2.5~m upstream of the target.  To quantify the effect of this
non-point-like neutrino production, we project the neutrino flux onto
a 20~m sphere centered on the Hg target and determine an effective
production angle based on the neutrino's projected location.  In this
model, the total anisotropy of the SNS neutrino flux 20~m from the
target center is $\sim$~5\%.  The dominant contribution is an excess
near $\cos\theta \approx -1$ consistent with neutrino production
within the PBW, with a secondary excess near $\cos\theta \approx 1$
consistent with neutrinos produced by decays in flight.  For a small
detector at the COH-CsI location 19.3~m from the target center and at
$\cos\theta\approx 0$, we predict less than a 1\% deficit of the
neutrino flux compared to the isotropic point-source approximation.
The contributions to the neutrino flux error from geometric
considerations are small, and add negligibly in quadrature to the 10\%
to the overall neutrino flux incident on our detectors in Neutrino
Alley.  The anisotropy depends on the relative contributions of the
different materials in our SNS geometry outlined in Table
\ref{tab:dims}, and emphasizes the need for new pion-production
measurements such as those discussed in Section \ref{sec:future}.

\section{Neutrinos at the Second Target Station}
\label{sec:sts}

\begin{figure}
  \centering
  \includegraphics[width = \columnwidth]{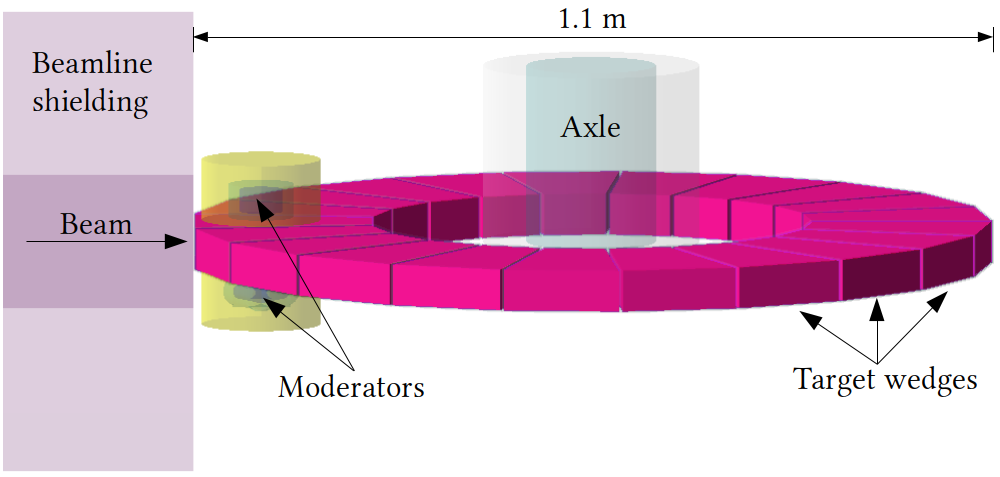}
  \caption{Geant4 implementation of the Second Target Station target
    and moderators.}
  \label{fig:stsGeom}
\end{figure}

We also created a model geometry to estimate the neutrino production
at ORNL's planned Second Target Station (STS) \cite{stsTDR}.  With a
projected completion in the early 2030s, COHERENT is engaged with the
design phase of this facility to optimize location and shielding with
the aim to deploy 10-ton-scale detectors for CEvNS and other physics.
Using preliminary details about the planned target provided at the
Workshop on Fundamental Physics at the Second Target Station in 2019
\cite{fpsts19}, we modeled 21 tungsten wedges surrounded by thin
layers of tantalum and water, evenly spaced in an assembly with a
1.1~m diameter.  We also modeled neutron moderators above and below
the active target wedge, centered along the beamline axis.  We
simulated a 6~cm (width) $\times$ 5~cm (height) beam profile to ensure
that the profile is smaller than that of a single tungsten wedge and
included the aluminum PBW and beamline shielding as implemented in our
First Target Station (FTS) geometry.  This target geometry is
illustrated in Fig.~\ref{fig:stsGeom} and is centered inside a 5~m
vaccuum box, then enclosed in a steel box (10~m outer edge, 5~m inner)
to mimic pion production in typical shielding materials without
assuming the geometry of the STS target surroundings.

\par With this simple geometry and 1.3~GeV incident protons, our
simulations predict 0.13 $\nu_X$ / POT for $\nu_\mu, \bar{\nu}_\mu$,
and $\nu_e$ from the $\pi^+$ decay chain, resulting in an approximate
total 0.39 $\nu$ / POT.  This estimate is larger than the predictions
for the FTS operating at 1.3~GeV due to the increased density of a
solid tungsten target.  We cannot accurately discuss the decay-at-rest
fraction of neutrinos or relative impact of the PBW since the
shielding surrounding the target remains unknown.  However, we note
that protons do escape the end of the 25-cm thick active target wedge
with enough energy to produce pions downstream of the target.

\par The STS will receive one of every four pulses from the SNS linear
accelerator and will operate as a 15~Hz, 0.8~MW facility.  Possible
locations for 10-ton scale COHERENT detectors at the Second Target
Station have been identified within a few tens of meters from the
planned target location.  With a tungsten target rather than mercury,
hadron production experiments using a range of targets at lower beam
energies will be useful in benchmarking our predictions \cite{loi}.

\section{Light Dark Matter Production at the Spallation Neutron Source}
\label{sec:darkmatter}

This work was focused on understanding the neutrino fluxes but also
explored the creation of other interesting particles.  In particular,
$\pi^0$, $\eta^0$, and $\pi^-$ production are relevant to dark matter
searches using the SNS as an accelerator~\cite{Dutta:2020vop,
  PhysRevD.102.052007}.  Here, we present some findings regarding the
production of such particles using QGSP\_\hspace{2pt}BERT, noting that
no effort was made in this work to specifically validate the
production of any hadrons other than $\pi^+$.  As mentioned in
Section~\ref{sec:physlists}, $\eta$ production is excluded from our
discussion here because it is not predicted by QGSP\_\hspace{2pt}BERT.
Predictions with QGSP\_\hspace{2pt}BIC have previously been used with
this simulation geometry to predict $\eta$ flux for sensitivity
studies~\cite{PhysRevD.102.052007}.

\begin{figure}
  \includegraphics[width = \columnwidth]{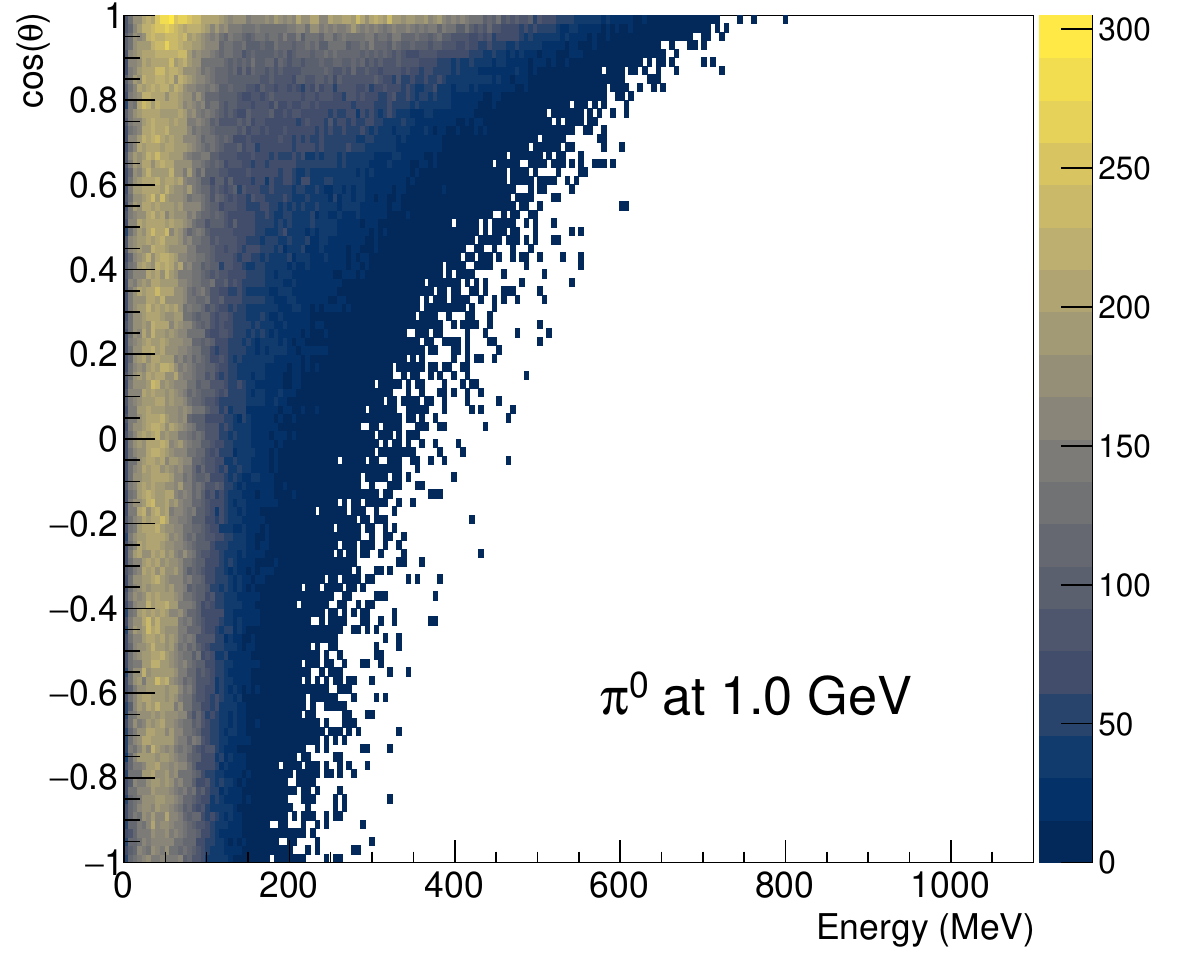}
  \includegraphics[width = \columnwidth]{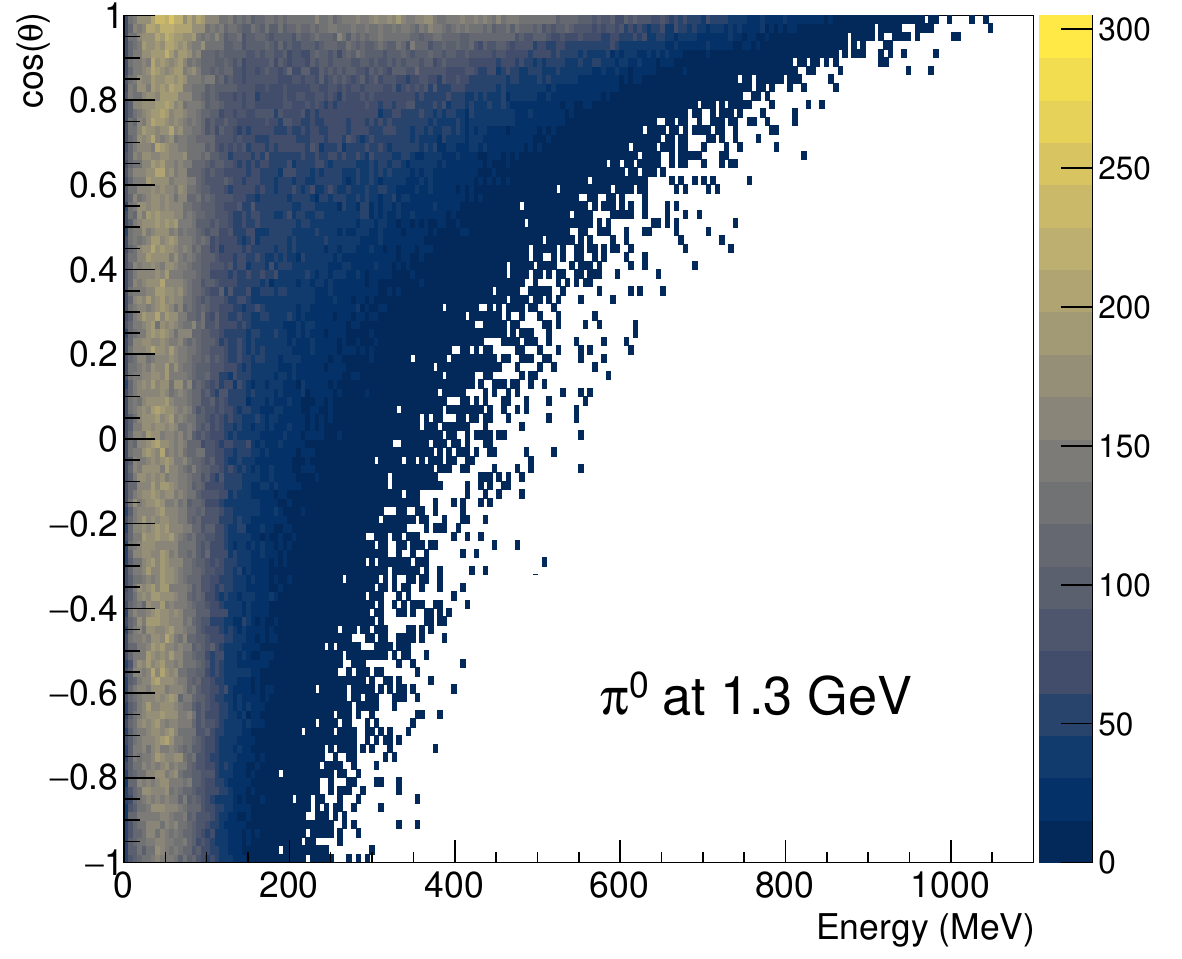}
  \caption{Distribution of production angles and creation energies of
    $\pi^0$ at the SNS-FTS assuming an aluminum PBW.  Top: Using
    1~GeV incident protons to mimic the current operating conditions
    of the SNS.  Bottom: Using 1.3~GeV incident protons to mimic the
    operating conditions following the upgrade.}
  \label{fig:dmAngles}
\end{figure}

Figure \ref{fig:dmAngles} shows the scattering angle as it relates to
the creation energy for SNS-produced $\pi^0$ from 1~GeV incident
protons on the top and 1.3~GeV incident protons on the bottom.  We
observe strong forward production for both, but note that we will have
a small flux directed towards Neutrino Alley (cos$ \theta \approx $0).
This is relevant primarily for the $\pi^0$ which could decay in flight
into dark-matter particles that cause an observable nuclear recoil in
our CEvNS detectors.  For $\pi^-$, dark matter could be produced in an
absorption process or in a charge-exchange process; both are more
efficient at non-relativistic energies, and each would emit particles
isotropically and negate any impact of forward production.

Assuming an aluminum PBW, the SNS produces 0.11 $\pi^0$/POT and 0.05
$\pi^-$/POT for 1~GeV incident protons.  We also predict that the
upgraded 1.3~GeV incident protons will produce 0.17 $\pi^0$/POT and
0.09 $\pi^-$/POT.  This study also demonstrates the potential gain of
the STS for dark matter searches, particularly in aiming for
forward-positioned detectors that reduce the distance to the target.

\section{Ongoing Efforts}
\label{sec:future}

\begin{figure}
  \includegraphics[width = \columnwidth]{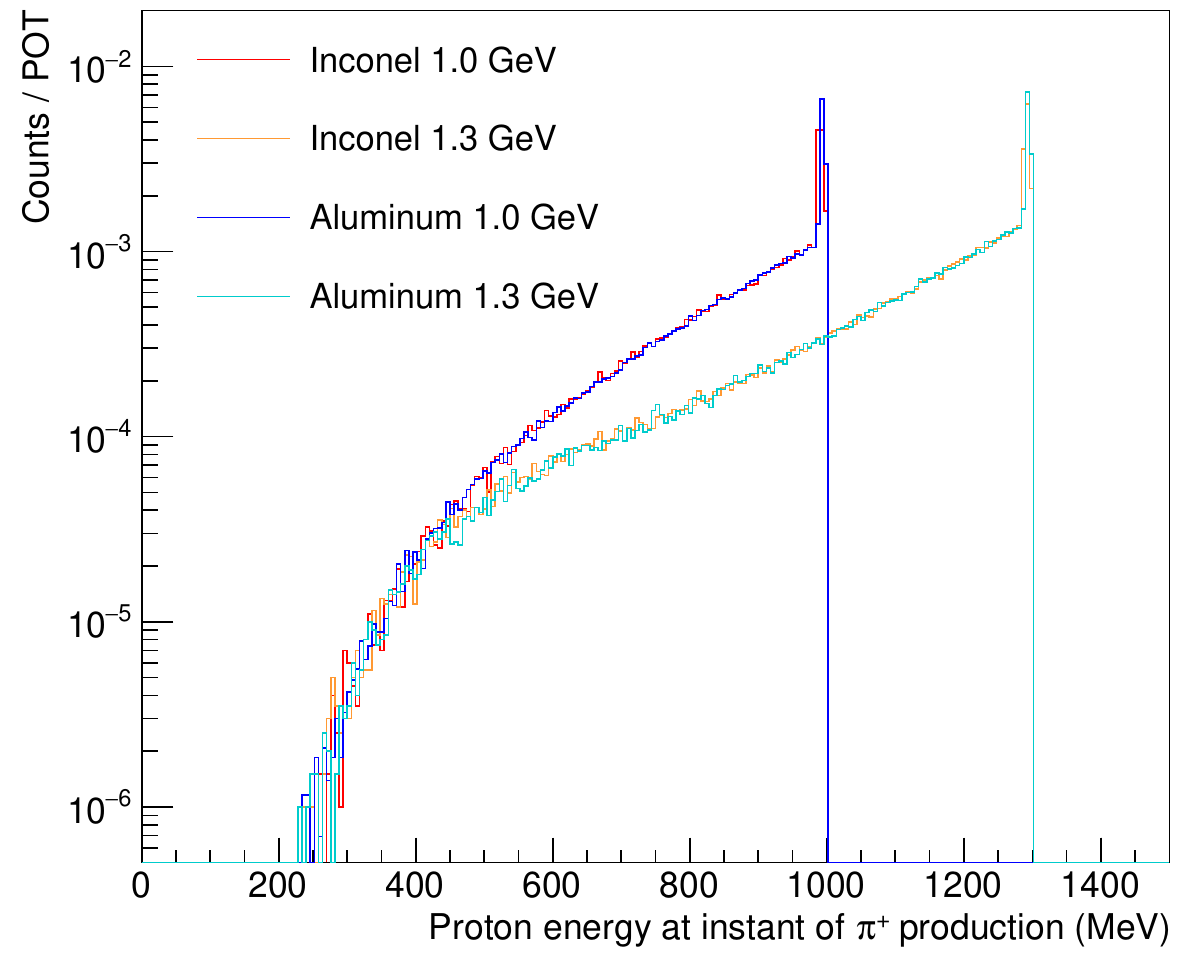}
  \includegraphics[width = \columnwidth]{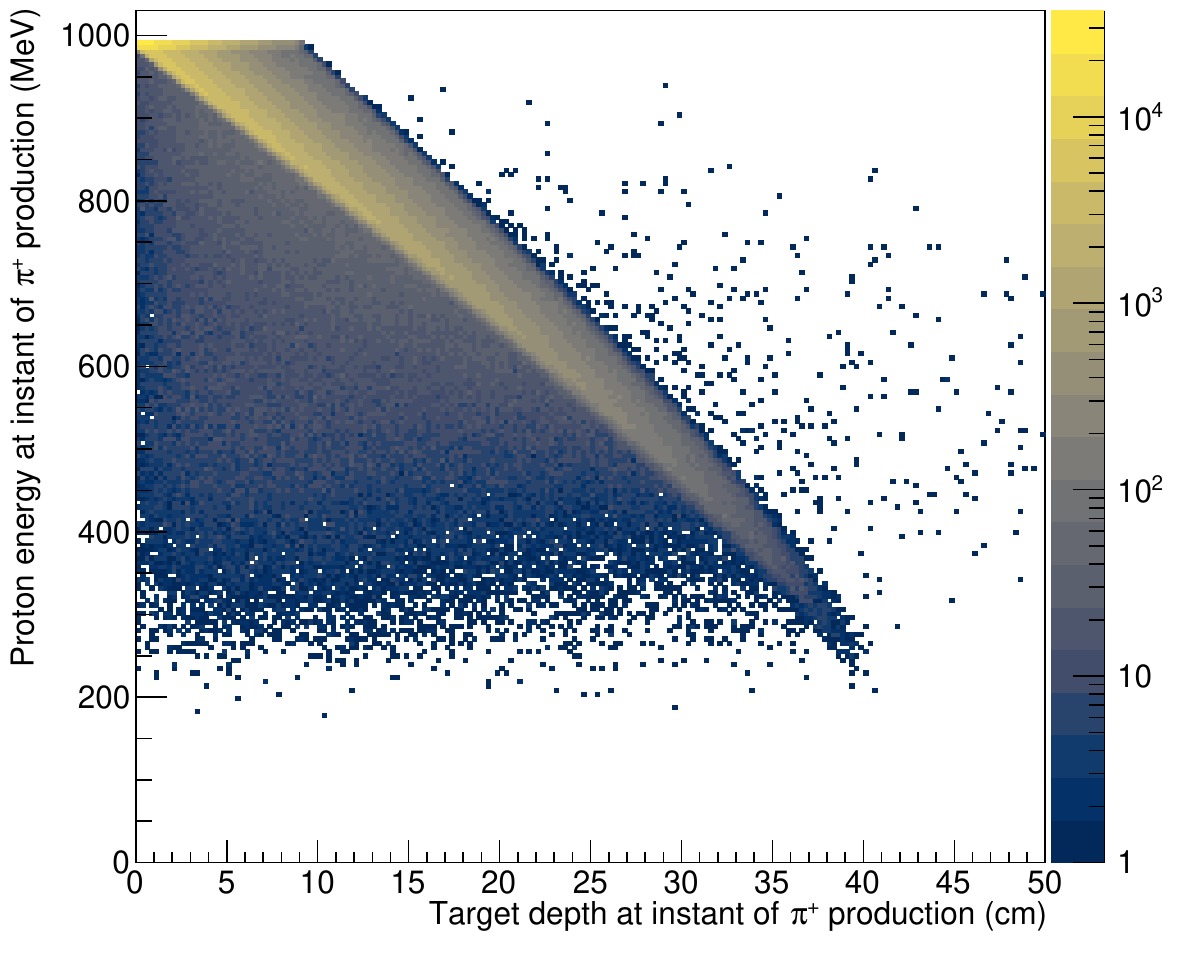}
  \caption{Top: A histogram of the proton energies which produce
    $\pi^+$ at the SNS.  Bottom: A closer look at how protons lose
    energy in the Hg target before creating $\pi^+$.}
  \label{fig:pe}
\end{figure}

In the absence of either pion-production data for protons incident on
Hg at energies up to 1.3~GeV or a precise measurement of the proton
energy-loss profile within the SNS target, the $\sim$~10\% uncertainty
assigned to our neutrino flux is a robust estimate that cannot be
significantly improved through simulation. Two types of experimental
measurements could further reduce this uncertainty.

Pion-production measurements with thin Hg targets would allow us to
validate our simulation against interactions on the same material as
the SNS target. The proposed EMPHATIC experiment at the Fermilab Test
Beam Facility could measure differential pion-production cross
sections on Hg at proton energies as low as 2~GeV with expected
uncertainties less than 10\%~\cite{emphatic:2019}. Meanwhile, the
NA61/SHINE collaboration~\cite{na61shine:2014}, which has measured
pion production on both thin and replica targets for a variety of
accelerator neutrino experiments, is investigating the possibility of
reducing the energy of the CERN SPS H2 proton beamline to 1~GeV for
low-energy pion-production studies~\cite{na61shine:2020}. These
measurements will benefit neutrino experiments at the SNS and at other
pion decay-at-rest neutrino sources with GeV-scale protons incident on
a mercury target such as the JSNS$^2$ sterile-neutrino search at the
Japan Spallation Neutron Source~\cite{JSNS-TDR:2017}.  As a
decay-at-rest source is insensitive to the production angle of the
pion, new pion-production measurements should ideally cover as close
to a $4\pi$ acceptance as possible.

Thin-target data at $\geq 1$~GeV, however, cannot account for the
effects of proton energy loss in the SNS target and from scattering in
the PBW as shown in Fig.~\ref{fig:pe}.  A separate approach to
reducing neutrino flux uncertainties would directly measure the total
neutrino production at the SNS target. A D$_2$O detector, deployed at
the SNS, would measure the charged-current interaction

\begin{equation}
\nu_e + d \rightarrow p + p + e^-.
\end{equation}

The cross section of this reaction is well understood; theoretical
calculations, taking several disparate approaches, have converged to
the 2--3\% level~\cite{mosconi:2007, Ando:2020}. A moderately sized
detector, about 680~kg, could achieve similar statistical precision in
about four SNS beam-years of operation. The observed $\nu_e$ flux from
the SNS target could then be multiplied by three to obtain the total
flux of all three neutrino flavors generated by $\pi^+$ decay. The
COHERENT collaboration plans to build such a detector to directly
normalize the simulated SNS neutrino flux~\cite{d2o-concept}.  We note
that if the neutrino flux can be independently measured to high
precision, one can in principle use neutrino data to validate models
of hadron and neutrino production.

\section{Conclusions}
\label{sec:theEnd}
Using Geant4.10.06's standard QGSP\_\hspace{1pt}BERT physics list and
treating the SNS as a point source, we predict a neutrino flux of 4.7
$\times$ 10$^7$ $\nu$ cm$^{-2}$ s$^{-1}$ at 20 m from the target with
$\sim$~99\% of the total flux produced by the stopped $\pi^+$ decay
chain for 1~GeV incident protons at the 1.4~MW First Target Station.
Our calculation has a 10\% uncertainty on the underlying
pion-production model.  This shared systematic for all COHERENT
detectors is now the dominant systematic uncertainty on our CEvNS
measurements, along with statistics.

\par Our simulation remains an invaluable tool for estimating the flux
of various particles at the SNS, the dependence of our predictions on
the incident proton energy, and relative effects of the beamline
geometry.  In the future, we intend to use a modified version of this
simulation to predict the low-energy contribution to the SNS neutrino
flux from $\beta^\pm$ decays coincident with a proton spill resulting
from activated materials.  We also intend to use the framework we have
developed here to perform model validation studies for particles such
as $\pi^0$, $\pi^-$ and $\eta$ that are relevant for future dark
matter studies at the SNS.

\par\noindent\textbf{\\Acknowledgements\\}

The COHERENT collaboration acknowledges the resources generously
provided by the Spallation Neutron Source, a DOE Office of Science
User Facility operated by the Oak Ridge National Laboratory. This work
was supported by the US Department of Energy (DOE), Office of Science,
Office of High Energy Physics and Office of Nuclear Physics; the
National Science Foundation; the Consortium for Nonproliferation
Enabling Capabilities; the Institute for Basic Science (Korea, grant
no.\,IBS-R017-G1-2019-a00); the Ministry of Science and Higher
Education of the Russian Federation (Project ``Fundamental properties
of elementary particles and cosmology'' No. 0723-2020-0041); and the
US DOE Office of Science Graduate Student Research (SCGSR) program,
administered for DOE by the Oak Ridge Institute for Science and
Education which is in turn managed by Oak Ridge Associated
Universities. The authors would also like to thank the Undergraduate
Research Office of Carnegie Mellon University for their support on
this project.  Sandia National Laboratories is a multi-mission
laboratory managed and operated by National Technology and Engineering
Solutions of Sandia LLC, a wholly owned subsidiary of Honeywell
International Inc., for the U.S. Department of Energy's National
Nuclear Security Administration under contract DE-NA0003525. The
Triangle Universities Nuclear Laboratory is supported by the
U.S. Department of Energy under grant DE-FG02-97ER41033. Laboratory
Directed Research and Development funds from Oak Ridge National
Laboratory also supported this project. This research used the Oak
Ridge Leadership Computing Facility, which is a DOE Office of Science
User Facility.

\bibliography{nuFluxPaper}

\end{document}